\documentclass[a4paper,fleqn,usenatbib]{mnras}

\pdfoutput=1

\usepackage{newtxtext,newtxmath}

\usepackage[T1]{fontenc}
\usepackage{ae,aecompl}

\usepackage{graphicx}	
\usepackage{amssymb}	
\usepackage{subfig} 

\title[Outer Boundary Condition in AGB stars]{Influence of the Outer Boundary Condition on models of AGB stars}

\author[G. Wagstaff et al.]{
G. Wagstaff,$^{1}$\thanks{E-mail: wagstaff@mpa-garching.mpg.de}
A. Weiss,$^{1}$
\\
$^{1}$Max-Planck-Institut f\"{u}r Astrophysik, Karl-Schwarzschild-Str. 1, 85748 Garching, Germany
\\
}

\date{Accepted 29.03.2018}

\pubyear{2015}

\begin{document}
\label{firstpage}
\pagerange{\pageref{firstpage}--\pageref{lastpage}}
\maketitle

\begin{abstract}
Current implementations of the stellar atmosphere typically derive boundary conditions for the interior model from either grey plane-parallel atmospheres or scaled solar atmospheres, neither of which can be considered to have appropriate underlying assumptions for the Thermally Pulsing Asymptotic Giant Branch (TP-AGB). This paper discusses the treatment and influence of the outer boundary condition within stellar evolution codes, and the resulting effects on the AGB evolution. The complex interaction of processes, such as the third dredge up and mass loss, governing the TP-AGB can be affected by varying the treatment of this boundary condition. Presented here are the results from altering the geometry, opacities and the implementation of a grid of MARCS/COMARCS model atmospheres in order to improve this treatment. Although there are changes in the TP-AGB evolution, observable quantities, such as the final core mass, are not significantly altered as a result of the change of atmospheric treatment. During the course of the investigation, a previously unseen phenomena in the AGB models was observed and further investigated. This is believed to be physical, although arising from specific conditions which make its presence unlikely. If it were present in stars, this phenomenon would increase the carbon-star lifetime above 10Myr and increase the final core mass by $\sim0.1M_{\odot}$ in the narrow initial-mass range where it was observed ($\sim2-2.3M_{\odot}$).
\end{abstract}






\section{Introduction}

The outer boundary condition (BC) in stellar evolution codes is a necessity arising from the numerical and physical difficulties which one would encounter if trying to integrate directly from a naive $P=0$ point outside of the star, right through to the centre.

The interior models of stellar evolution calculations solve the equations of stellar evolution and typically take a BC from sources such as plane-parallel calculations with grey opacities, or scaled solar atmospheres. Work has been done to investigate the effect of changing this treatment ~\citep{VandenBerg2008} however, this typically concentrates on earlier evolutionary phases. 

The Thermally Pulsing Asymptotic Giant Branch (TP-AGB) is both a highly complicated and hugely important phase in stellar evolution, not just for the stars themselves but for the surrounding environment and galactic chemical evolution.

During this double shell burning phase, unstable burning in the thin He-burning shell causes thermal pulses which dramatically impact on the star, and may result in a 3rd dredge up (3DU) event. This is due to the convective envelope briefly extending into regions which have previously undergone nuclear burning, changing the chemical composition at the surface. This mechanism is responsible for several effects which give AGB stars added significance in a wider context, particularly when coupled with the high mass-loss rates which occur during this period. For instance, this is the process by which carbon stars are produced.

Furthermore, the nuclear processes which occur in the interior of TP-AGB stars extends beyond simple burning of hydrogen and helium, as it is thought this is the main site for s-process nucleosynthesis, the source of half of the elements heavier than iron \cite{Arlandini1999}. Additionally, higher mass AGB stars can undergo a process known as hot bottom burning (HBB), during which the base of the convective envelope overlaps with part of the H-burning shell, resulting in efficient conversion of dredged-up carbon into nitrogen. This can delay the formation of carbon stars.

To evolve into the central stars of planetary nebulae, AGB stars must expel the majority of their convective envelope, which is done during the TP-AGB phase with significantly increased mass loss, reaching rates as high as $\dot{M}_{\rm AGB} \sim 10^{-6} - 10^{-4} M_{\odot} {\rm yr}^{-1}$ \citep{Wood1992}. Since early attempts at incorporating so called superwind prescriptions into evolutionary modeling \citep{Vassiliadis1993,Bloecker1995} both theoretical models \citep{Wachter2002} and observations \citep{vanLoon2005, Goldman2017} have tried to improve on the description, and understand the differences between O- and C-rich stars. This is a hugely important aspect of TP-AGB modeling, as this determines when the phase ends, and the composition of material ejected into the ISM.

The importance of calculating AGB models extends beyond the understanding of the stars themselves, with the results being important for galactic chemical evolution \citep{Kobayashi2011} and determining the transition between which stars form planetary nebulae, and which go onto forming electron capture supernovae \citep{Doherty2015}. In addition, these stars provide commonly accepted explanations for the multiple populations in Globular Clusters \citep[for a review, see][]{Gratton2012} and the abundance pattern in CEMP-s stars \citep{Abate2016}.

AGB models have been produced for a variety of reasons, all with the common goal of understanding this crucial phase of stellar evolution. Some tend to focus on the nucleosynthesis \citep{Karakas2007,Cristallo2009,Cristallo2011} while synthetic codes avoid full model calculations, and are more suitable for population synthesis (e.g. \cite{Marigo2013}). All calculations tend to rely on similar methods of treating the outer BC, although the TP-AGB is a phase in which the typical underlying assumptions are likely to break down. This could potentially play a role in the evolution which is not as readily expected to manifest itself during the main sequence evolution.

Indeed, it is possible to show through approximate analytic solutions \citep[see][Ch. 11.3.3]{Kippenhahn2012} that for stars with a radiative envelope, the interior solution rapidly converges to the same solution almost independent of the boundary condition provided. However, based on the same simplifying assumptions, it can be shown that for stars with a convective envelope, changes in the outer BC influence the interior solution to a larger extent.

In order to determine whether this is the case, this work investigates the effect of altering the treatment of the stellar atmosphere in models calculated with the Garching stellar evolution code (GARSTEC, \cite{Weiss2008}) following previous work on the TP-AGB in \cite{Kitsikis2008}. Assumptions of geometry and opacities are considered, along with grids of MARCS \citep{Gustafsson2008} or COMARCS \citep{Aringer2009} model atmospheres. 

To begin with, the following section, \S \ref{Sec: SE code/}, outlines the theoretical framework of the atmospheric calculation with respect to the stellar evolution code. \S \ref{Sec: Calculations} discusses the calculations preformed during the investigation, and the reasoning behind any choices made. The primary results are then considered in \S \ref{Sec: Results}, with particular attention paid to the previously unseen behaviour in \S \ref{Anom}, before being discussed more broadly along with future work in \S \ref{Sec: Discussion}.


\section{Details of the Stellar Evolution Code}

\label{Sec: SE code/}

The stellar evolution code GARSTEC, as described in ~\cite{Weiss2008} with AGB specific alterations outlined in ~\cite{Weiss2009}, was used for all calculations. Calculations presented here are consistent with the physics described in the above papers, and all details of the applied physics can be found therein. As the intention of this paper is to compare the influence of changing the atmospheric treatment, it is the implementation of the boundary condition which is primarily discussed within this paper.

Although not altered for this work, it is worth drawing attention to the mass loss prescriptions implemented on the TP-AGB, given their significant influence on the evolution and calculation results. For the case of O-rich stars, the prescription of ~\cite{vanLoon2005}
\begin{eqnarray}
  \log \dot{M} = &-& 5.65 + 1.05\cdot \log\left( 10^{-4}\frac{L}{L_{\odot}}\right) \nonumber \\
    &-& 6.3\cdot \log\left( \frac{T_{\rm eff}}{3500K}\right)
\end{eqnarray}
and of ~\cite{Wachter2002} for C-rich stars
\begin{eqnarray}
  \log \dot{M} = &-&4.52 +2.47\cdot \log\left( 10^{-4}\frac{L}{L_{\odot}}\right) \nonumber \\
  &-& 6.81\cdot \log\left( \frac{T_{\rm eff}}{2600K}\right) - 1.95\cdot log\left( \frac{M}{M_{\odot}}\right)
\end{eqnarray}

is taken for the TP-AGB mass loss.

It is worth noting that although large uncertainties remain in mass loss prescriptions, or the $T_{\rm eff}$ which they refer to, a consistent picture emerges amongst all relations of this high $T_{\rm eff}$ dependence, typically $\dot{M} \propto T_{\rm eff}^{a}$, where $a=6...7$. This has the potential to be significantly influenced by the outer BC, which is itself closely connected to the $T_{\rm eff}$. 

As implemented for this work, these relations only come into effect when the pulsation period exceeds 400 days, as determined by the relation
\begin{eqnarray}
\log(P_0/d) = -1.92 -0.73\log M + 1.86\log R
\end{eqnarray}
relating the period to the mass and radius of the star, taken from ~\cite{Ostlie1986}. This is done to limit the application of the strong mass-loss prescription to models similar to the observed stars where the mass-loss rates were determined. When $P<400\rm d$, the mass loss reverts to a Reimers rate \citep{Reimers1975}. This cutoff period is in fact intended for the O-rich case only, as the relation used there is derived from observations of such stars, however, it is also applied in the case of the C-stars, where a theoretical prescription is used.

\subsection{Grey Plane Parallel Atmosphere}

The use of plane-parallel grey atmospheres in Local Thermodynamic Equilibrium (LTE), as first outlined by Eddington ~\citet[p320]{Eddington1959}, is widespread in the stellar evolution community. It allows for a quick and reliable method of generating a BC unique for a star's position within the HR diagram, based on particular simplifications. 

The geometric simplification involves considering certain parameters (radius, luminosity, interior mass) to be constant throughout the atmosphere, which allows the equation of hydrostatic equilibrium to be combined with the definition of optical depth ($d\tau/dr = \kappa\rho$) to obtain
\begin{eqnarray}
\label{eq: DtauDp}
 \frac{d\tau}{dP} = g\kappa
\end{eqnarray}
with the surface gravity $g=GM/R^2$ for constant stellar mass, M, and radius, R. 

The temperature stratification is then taken from the derivation by Eddington under the assumptions of plane-parallel, grey atmospheres such that
\begin{eqnarray}
 T^4 = \frac{3}{4} T_{\rm eff}^4 \left[\tau + \frac{2}{3} \right] 
\end{eqnarray}
in the range $\tau=[0,2/3]$. However, this can be generalised by replacing the factor 2/3 in the brackets with the Hopf function $q(\tau)$, which allows scaled solar relations to be implemented such as that of \cite{KrishnaSwamy1966}.

Equation \ref{eq: DtauDp} is integrated from $\tau=0$, with $P(\tau=0)=0$, to $\tau=2/3$ using a Adams-Bashford/Adams-Moulton predictor/corrector method to obtain the pressure at the outermost point of the interior model. In conjunction with the Stefan-Boltzmann law ($L=4\pi R^2\sigma T_{\rm eff}^4$), this provides the outer boundary condition for the stellar evolution code.

In this instance no mass is included in the atmosphere, therefore $M(\tau=2/3)=M_*$. However, in the following spherically symmetric approach it is possible to account for the mass in the region where $\tau<2/3$, which at its maximum reaches $\sim 10^{-3}M_*$ for this analytic treatment.

\subsection{Spherical Geometry}

The assumption that the stellar atmosphere can be approximated as plane-parallel may be valid on the Main Sequence, it is however hard to justify on the RGB and AGB where the atmospheric extent is expected to be a significant proportion of the stellar radius. In order to investigate this assumption a spherical grey atmosphere, as outlined in ~\cite{Lucy1976}, was implemented. 

Within this framework, the effective temperature is taken analogously with the Eddington solution, i.e. $T_{\rm eff} = T_*$ at $\tau = 2/3$, where $T_*$ is referred to as the photospheric temperature to acknowledge the lack of a unique definition of $T_{\rm eff}$ in an extended atmosphere. This approach has the advantage of reducing to the Eddington plane-parallel atmosphere already implemented within GARSTEC, when the geometric extent of the atmosphere is negligible.  This results in a consistent definition and implementation of $T_{\rm eff}$, allowing the influence due to geometry to be isolated.

The temperature stratification is given by
\begin{eqnarray}
 T^4 = \frac{1}{2}T^4_*[2W  +  \frac{2}{3}\tau]
\end{eqnarray}
where W is the geometric dilution factor $W=\frac{1}{2}\{1-\sqrt{[1-(R/r)^2]}\}$ and the optical depth is defined by
\begin{eqnarray}
\label{eq: DtauDr}
 \frac{d\tau}{dr} = - \kappa r \left(\frac{R}{r}\right)^2
\end{eqnarray}

In order to implement the spherical approximation, equation \ref{eq: DtauDr} is integrated along with the equation of hydrostatic equilibrium and making use of the equation of state. This is done using the same method as in the plane parallel case. However, an initial estimate of the outer radius is required to start the calculation in the spherical calculation. This results in an optical depth other than $\tau=2/3$ being reached when the radius equals the fitting radius. It is therefore necessary to estimate a new outer radius based upon the previous integration and then this process can be iterated until the fitting radius and optical depth both match within a given tolerance.

\subsection{Opacities}

To avoid time consuming and complicated radiative transfer calculations some definition of a frequency independent, or grey, opacity is required. Under the assumption of Local Thermodynamic Equilibrium (LTE) and the Diffusion Approximation (DA), which are strictly valid only in the stellar interior, it is possible to derive the Rosseland Mean (RM) opacity ~\citep{Rosseland1924}, denoted by $\kappa_{\rm R}$ and related to the Planck function $B_{\nu}$ by the equation
\begin{eqnarray}
 \frac{1}{\kappa_{\rm R}(\rho,T)} = \frac{\int_0^{\infty} \frac{1}{\kappa (\nu)}\frac{\partial B_{\nu}}{\partial r}d\nu}
 {\int_0^{\infty}\frac{\partial B_{\nu}}{\partial T}d\nu}
\end{eqnarray}
at a particular density, $\rho$, and temperature, T, for a given chemical composition which results in the frequency dependent opacity $\kappa (\nu)$. 

The typical optical depth taken to be the surface of a star, $\tau =2/3$, corresponds to the depth from which photons experience on average less than a single scattering before escaping the star. This would therefore be the first point at which it could, with at least some justification, be claimed that the local conditions satisfy the DA, although in reality, it is likely to be even deeper, perhaps at $\tau=10$ although it is difficult to say exactly.

As such, in the outer layers of a star, the DA, which is a necessary condition for the RM opacity, breaks down as the mean free path is no longer much smaller than the depth. Therefore, an alternative is desirable. Unfortunately there is no analytically derived mean opacity for the optically thin stellar atmosphere, though one possibility is to use a straight average, or the Planck mean (PM) as outlined in ~\cite{Eddington1922}, such that
\begin{eqnarray}
 \kappa_{\rm P}(\rho,T) = \frac{\int_0^{\infty} \kappa(\nu) B_{\nu} d\nu}{\int_0^{\infty} B_{\nu}d\nu}
\end{eqnarray}

Tables of the PM molecular opacities at low temperatures were produced with the program described in ~\cite{Ferguson2005} and are equivalent to those produced in the same manner as for the RM opacities, although only for the solar C/O value. These tables are combined with OPAL-tables \citep{Iglesias1996} for high temperature opacities in the same way as has previously been done for the RM, low temperature tables as described in \cite{Weiss2008}.

The RM tends to favour opacity intervals in the spectrum, which allow the majority of the radiation flux to pass through at frequencies of least resistance. This is likely to result in a systematic undersampling of the actual mean opacity in a stellar atmosphere. On the other hand, as the PM opacities are a straight average, the high (but spectrally narrow) line and molecular opacities which become particularly important in the outer layers of AGB stars can be expected to artificially raise the average opacity.

Given the expectation that one treatment undersamples the opacity, while the other oversamples it, the two should result in an upper and lower bound for what is possible in the context of a grey atmospheric treatment.

As for implementation, it is well accepted \cite[p376]{Hubeny2014} that the RM opacity is suitable for stellar evolution models within the interior, where the diffusion approximation holds. If we also take this to be true, at least predominantly, at $\tau = \frac{2}{3}$, then it is desirable to recover the limit of the RM at the lower boundary of the atmospheric calculation. 

Conversely, at $\tau = 0$, the justification for using the RM is at its weakest, lines are expected to have a greater influence on the opacities and as such this is where the PM is, if not reasonable, at least worth considering as an extreme case to determine if the use of RM opacities in unjustified regions is having a substantial influence on the atmospheric treatments, and subsequently the stellar evolution models.

Given this reasoning, the following method for the opacity treatment within both the plane-parallel and spherically-symmetric atmospheric calculations was implemented in GARSTEC. A transition for the opacity as a function of the optical depth in the range $\tau=[0,2/3]$ is applied as follows
\begin{eqnarray}
\ln(\kappa(\tau)) = \ln(\kappa_{\rm P}(\tau)) +  \left(\ln(\kappa_{\rm R}(\tau)) - \ln(\kappa_{\rm P}(\tau))\right)*\left(\frac{\tau}{2/3}\right)
\end{eqnarray}
This function allows for the smooth change between the two mean opacity treatments and is implemented as it provides the desired treatment of PM at $\tau=0$ and RM at $\tau=2/3$ while providing at least a reasonable agreement with the models calculated using the COMARCS radiative transfer atmospheres, as is discussed further in section \ref{Sec: Plank Mean Proxy}.

Suffice it to say however that a grey atmosphere, with any mean opacity, will not be able to reproduce the detail of full radiative transfer calculations, although changing the mean opacity which is used should allow for something to be said about the range of results which are obtainable when using the typical grey atmospheric methods.

\subsubsection{$C/O$ ratio}

The C/O ratio is known to be hugely important in contributing to the opacities in the low temperature regimes experienced in the outer layers of AGB stars \citep{Marigo2002,Marigo2009}. There can be drastic changes due to the molecular lines which can form, with the underlying O-rich or C-rich abundance entirely dominating given the CO molecule has a high binding energy and freezes out, leaving the respective excess to govern additional molecule formation.

The dependence on the C/O ratio is investigated here, based on the implementation discussed in ~\cite{Weiss2009}. The RM opacities are taken from interpolation within the usual values at different C/O values, with an additional interpolation performed to reach the desired C/O value. This is done for all regions above the H-burning shell, including the atmospheric calculation, for the following values of C/O: 0.48 (solar), 0.9, 1.0, 1.1, 3.0 and 20. The PM opacities are only implemented at the solar C/O value, and are otherwise equivalent to the low temperature RM tables as described in ~\cite{Weiss2009}.

\subsection{Radiative Transfer}

\label{Sec: RT}

The previously mentioned cases are intended as a step to determine if the exterior boundary condition of stellar models influences the evolution of AGB stars, and which also allow particular assumptions to be investigated. However, in the case that this outer BC alters the evolutionary behaviour it is clear that a numerical solution to the equations of radiative transfer would be preferable as a way of providing a better description of the underlying physics, as the above assumptions become redundant. 

In order to investigate the effect of using full radiative transfer models, a grid of atmospheres was implemented. The atmospheric models used are the MARCS grid \citep{Gustafsson2008} and for the low surface gravity regions, the COMARCS grid ~\citep{Aringer2009,Aringer2016}. The grids are generally labelled by the parameters ($T_{\rm eff}$, log g, [Fe/H]) with an additional mass parameter in the spherical models. All COMARCS models are spherical, while for MARCS plane parallel models are used when log g>3 and spherical otherwise. The COMARCS models also include C/O as a parameter (in the range C/O=[0.275,10]), however only models with a solar value of the ratio (C/O = 0.55) were used in evolutionary calculations due to a limited grid being available at other C/O values. As such, the COMARCS grid is analogous to the MARCS models in terms of parameter labels ($T_{\rm eff}$, log g, [Fe/H]).

\begin{figure}
  \includegraphics[width=.9\columnwidth]{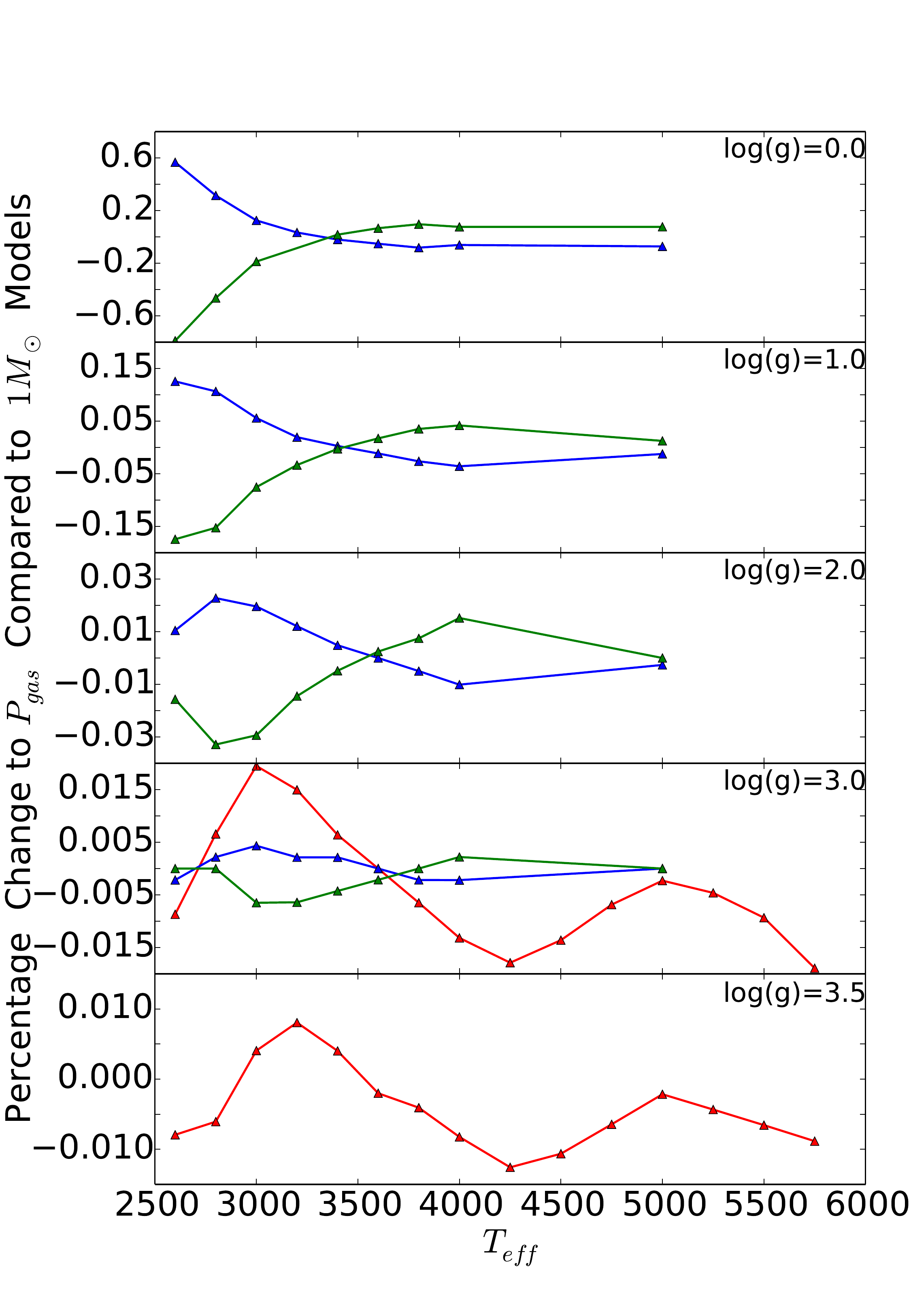}
   \caption{MARCS atmospheric models, percentage change in gas pressure at $\tau=1$ relative to a $1M_{\odot}$ model for red: plane-parallel, blue: 2$M_{\odot}$ and green: 0.5$M_{\odot}$ models across a range of $T_{\rm eff}$. The panels from top to bottom represent log g=[0,1,2,3,3.5], all models are solar metallicity.}
   \label{Fig: Pgas Diff MARCS}
\end{figure}

For lower surface gravities, spherically-symmetric models are produced instead of plane-parallel models, as the geometric extent of the more diffuse atmospheres becomes relevant. This results in an additional mass parameter, the dependence on which was investigated. However, the effect on the structural atmospheric quantities is minimal based on changes in the mass, whether plane-parallel models, or spherical models with masses ranging from $0.5-5M_{\odot}$, with the dependency on log g being far more important. To illustrate this, Fig \ref{Fig: Pgas Diff MARCS} shows the percentage change in the gas pressure at $\tau =1$, compared to the value for a $1M_{\odot}$ model for different MARCS atmospheres. Each panel represents the respective difference between the models at a given log g (log g=[0,1,2,3,3.5], across a range of temperatures. The blue line represents models with a mass of 2$M_{\odot}$, the green models with 0.5$M_{\odot}$ while the red shows the plane-parallel models. As the plane-parallel models are only available at the higher log g values (3, 3.5) they are only shown in the lower two panels, while there are no models for $0.5M_{\odot}$ or $2M_{\odot}$ at log g=3.5.

Clearly, in all cases, differences are limited to below 1\%, so it appears that the structural effect from the mass is not particularly significant. What differences do exist, are larger for the plane-parallel models than the differences which arise from changing the mass, at least at comparable stellar parameters. It should of course be noted that this is at the highest log g value (log g=3) available for the non solar mass models, which is when any effect arising from mass is expected to be a minimum, and the model should be close to the plane-parallel case. Of more relevance for this work, is the noticeable increase in the disparity for models in the low log g, low $T_{\rm eff}$ region, as can be seen in the top two panels of Fig. \ref{Fig: Pgas Diff MARCS}. Furthermore, the fact that the change from the $1M_{\odot}$ model to the 0.5 and 2$M_{\odot}$ models appear to always be in the opposite direction, would suggest that in principle it should be possible to interpolate the atmospheric properties within an additional mass parameter. However, as the grids of atmospheric models are, in both the MARCS and COMARCS cases, too sparse for any mass other than 1$M_{\odot}$, the boundary conditions provided by integration in the 1$M_{\odot}$ grid are used for all stellar models, with no explicit mass dependency (although this of course still enters into the log g parameter). 

The boundary condition is implemented by interpolation of the required structural values (pressure and temperature) within the grid parameters. This is typically only done  at the optical depth required for fitting to the interior model, however this can be done for all optical depth points in order to attach the full atmospheric structure (e.g. when a comparison is required). The interpolation is done in a piecewise manner, first interpolating in one of the grid parameters then another and so on, using a 4th order polynomial fit.

\section{Calculations}
\label{Sec: Calculations}

As the primary focus of this work is to study the influence of the outer BC on the TP-AGB evolution, models were separated into their evolution prior to the first TP, and the TP-AGB itself. This allows for a more consistent analysis of the models, as any change in the core mass at the onset of the first TP can itself become an influence, and in addition some of the assumptions being challenged by the investigation, such as the use of different opacities, may only be justifiable during this phase of a star's lifetime.

For completeness, models are also produced for the full evolution using different outer BCs , and discussed in section \ref{Sec: Results}, although this primarily illustrates that any influence is more pronounced on the TP-AGB. In this case it is stated the models are calculated from the main sequence using the stated BC, otherwise models using the plane-parallel atmosphere with Rosseland mean opacities up until the first TP are used. 

The models presented here are calculated with the solar abundance distribution of \cite{Grevesse2007} to match the MARCS grid, giving a solar metallicity of Z=0.012, lower than typically taken for stellar modeling. On the other hand the COMARCS atmospheres take the abundances from \cite{Anders1989} for all values other than C, N and O which are taken from \cite{Grevesse1994}. Although the metallicity taken in this work is lower than generally taken for solar composition models, the main purpose of this work is for comparison between atmospheric treatments, so it is not particularly significant. The difference in abundances assumed by the two radiative transfer grids, along with them being computationally different is another reason for separating the pre- and TP-AGB evolution, as there is a clear difference in the evolutionary models if a direct transition is attempted.

Given the various BC implementations and comparisons which are described hereafter, the following abbreviations are used:
\begin{itemize}
   \item pp: Plane-parallel
   \item ss: Spherically symmetric
   \item RM: Rosseland mean opacities
   \item PM: Planck mean opacities
   \item RM-CO: Rosseland mean opacities interpolated in C/O value
   \item RT: interpolation within MARCS (pre-1st TP) and COMARCS (TP-AGB) grid of atmospheres
   \item ms: model evolved from the main sequence with stated BC
\end{itemize}

Models implementing a RT grid would be the clear, physically motivated, preference for these calculations, however as problems were encountered with the low surface gravities reached during the TP-AGB evolution, there was a limit to how far such evolutionary models could be continued. As such, an alternative for modeling the outcome which could be reasonably expected from such a treatment was investigated, and is discussed in more detail in Sect. \ref{Sec: Plank Mean Proxy}

\subsection{C/O in COMARCS Atmospheres}

\begin{figure}
  \includegraphics[width=\columnwidth]{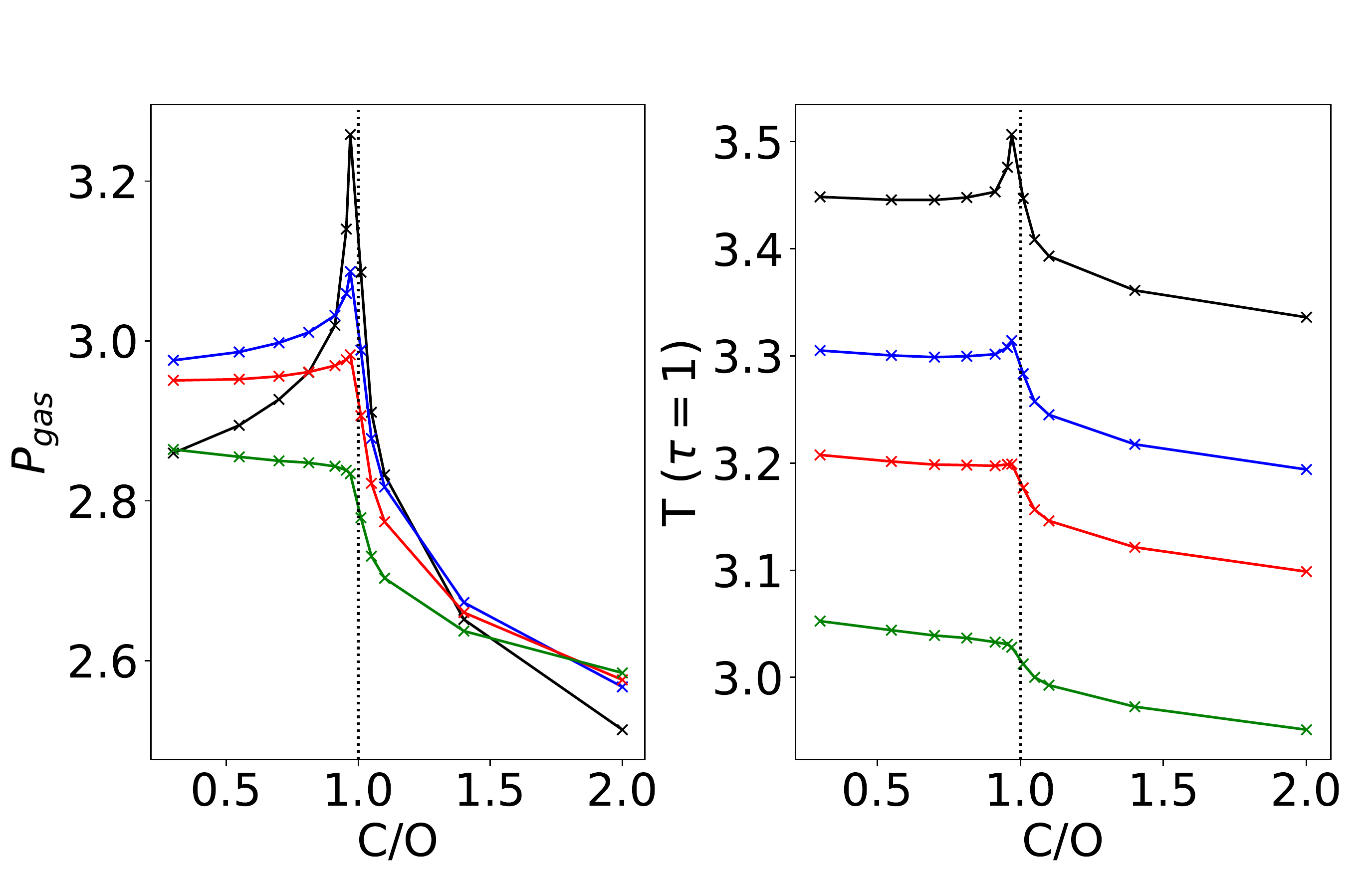}
  \caption{COMARCS atmospheric parameters ($P_{\rm gas}$, $T_{\rm gas}$) at $\tau=1$ for atmosphere at $\log g=0$, $\rm [Fe/H]=0$, $M=1M_{\odot}$ and $T_{\rm eff}=$ [2700K: black, 3000K: blue, 3200K: red, 3500K: green] for a varying carbon to oxygen ratio (C/O).}
  \label{Fig: COMARCS COvsT,P}
\end{figure}

It is already known that the opacities at a given metallicity vary as a function of changing C/O ~\citep{Marigo2002, Marigo2008}, and has been shown to influence the TP-AGB evolution ~\citep{Cristallo2007, Weiss2009} including at low metallicities \citep{Constantino2014}. It is nonetheless worth taking a moment to look directly at the COMARCS models which have at specific locations in the log g, $T_{\rm eff}$ grid a large number of detailed radiative transfer calculations with varying C/O. This results in a change in the atmospheric structure, producing a resulting change to the photospheric boundary condition taken for the stellar evolution code. 

Fig \ref{Fig: COMARCS COvsT,P} shows the gas pressure and structural temperature at $\tau=1$ for models with a different $T_{\rm eff}$ but the same log g and metallicity. The difference in these values, as a function of the C/O value, can be seen to vary quite drastically in the transition region between O-rich and C-rich, particularly at lower temperatures where molecules become increasingly important.

The increase in both pressure and temperature as C/O approaches 1, and subsequent decrease as the models become further carbon enhanced is a result of the CO molecule having such an overwhelming influence on the resulting molecular chemistry and the lack of it as all the C and O is bound up within the CO molecule. Unfortunately the extent of this grid is insufficient to interpolate in the C/O value during the evolutionary calculations and would require very careful consideration if done in the future, given the discontinuity in the slope at unity within the C/O value.

\subsection{Obtaining the Analysis}

There are inherent difficulties in analysing the often complex evolution during the TP-AGB, which regularly requires certain decisions to be taken about how a result is defined. As the objective during this work is to proceed in a manner which allows for direct comparison between two models which differ only in their treatment of the outer BC, how this is done can become fairly crucial, and as such taking a moment to outline how this is done is considered pertinent.

Of particular significance can be the method for obtaining final results during the evolution, with it often being the case that models cannot proceed through the end of the TP-AGB evolution, directly to the post-AGB phase due to issues of convergence. While it is possible to follow this evolution to the end of the star's nuclear burning lifetime, it usually requires some alteration of numerical parameters, and tends to be required on an individual model basis. In this investigation, this is not desirable. It is considered for the purposes outlined previously, better to use common methods of analysis of the models, rather than to force the evolution to complete the evolution.

Convergence has been known for many years to be a problem towards the end of TP-AGB evolution \citep{Wood1986}, encountered by various groups since its discovery \citep{Karakas2007,Weiss2009}, with the primary cause, a dominance of radiation pressure at the base of the convective envelope, more thoroughly investigated by \cite{Lau2012}. In general, it is easier to follow low to intermediate mass stars up until the last TP, although even then it is often the case that convergence becomes an issue as the star loses the remainder of its convective envelope, forming the central stars of planetary nebulae \cite{Bertolami2016}. Unfortunately, the exact moment during the final TP cycle where calculations stop is not always consistent, influencing the final core mass obtained due to the cycle of core growth from burning and reduction due to the 3DU. The definition of final core mass is thus taken to be the extent of the hydrogen-free core at the TP prior to the final TP.

A similar approach is taken in producing the yields for the models, removing the remainder of the envelope at the second to last TP, giving the overall material ejected by the star. Beyond this inclusion of the final envelope mass in the yields, no further extrapolation is used in any of the results presented here. As such any final values which are given for the more massive stars, and typically encounter convergence issues with multiple TP cycles remaining incomplete, should not be taken as definitive.


\section{Results}
\label{Sec: Results}

\subsection{Planck Mean as a Proxy}
\label{Sec: Plank Mean Proxy}

\begin{figure}
  \includegraphics[width=.9\columnwidth]{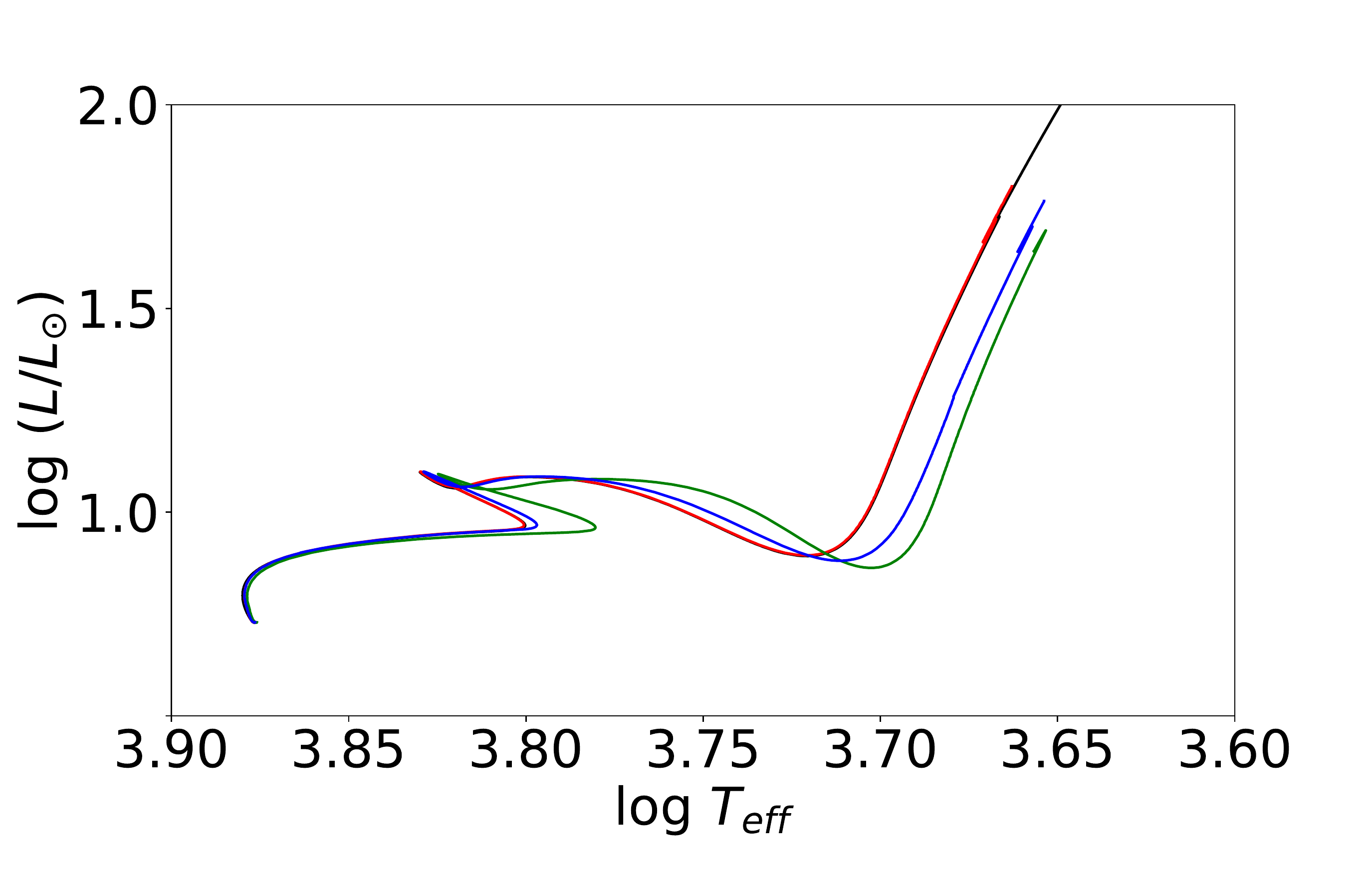}
   \caption{HR-diagram for $1.5M_{\odot}$ models with outer BC denoted by colour. Black: RM-pp, red: RM-sph, blue: RT (MARCS), green: PM-pp}
   \label{Fig: hrd_preAGB}
\end{figure}

Although it is not possible to use radiative transfer models to cover the entire TP-AGB evolution, it is still desirable to be able to consider what the overall influence may be if such a possibility were available. 

When viewing the different evolutionary tracks, the first thing which became apparent is the influence on the effective temperature of the models. Of particular relevance is that in both the cases of using RT models and the PM opacities, the effect is to move the models to a lower temperature. However this only occurs as the models move away from the main sequence evolution. Fig. \ref{Fig: hrd_preAGB} demonstrates this, showing the evolution of a $1.5M_{\odot}$ model with different atmospheric treatments. The evolutionary tracks overlap on the MS until the development of a convective core, only fully separating as the stars begin to ascend the RGB.

Although justification for using the PM opacities during the earlier evolution is perhaps even less than on the TP-AGB, Fig. \ref{Fig: hrd_preAGB} gives a clear indication that stellar evolution models are predominantly independent of the outer boundary condition during MS evolution, at the masses considered here. Concurrently, it shows that as the star expands, traversing the HR diagram into the red giant region, the effective temperature of the model is altered.

Additionally, the tracks suggest that the RT models are indeed bound by the two opacity treatments, and that on the RGB, more closely follow the PM models. A similar behaviour is observed on the TP-AGB for models using the COMARCS atmospheres, rather than the MARCS models used in this example. This leads to the interesting question of whether the PM could in some way be used to allow for the study of the stellar evolution models throughout the low log g environment, which was the intended subject of this study, but where it is computationally difficult to produce sufficient RT models to allow for a direct investigation.

\begin{figure}
  \includegraphics[width=.9\columnwidth]{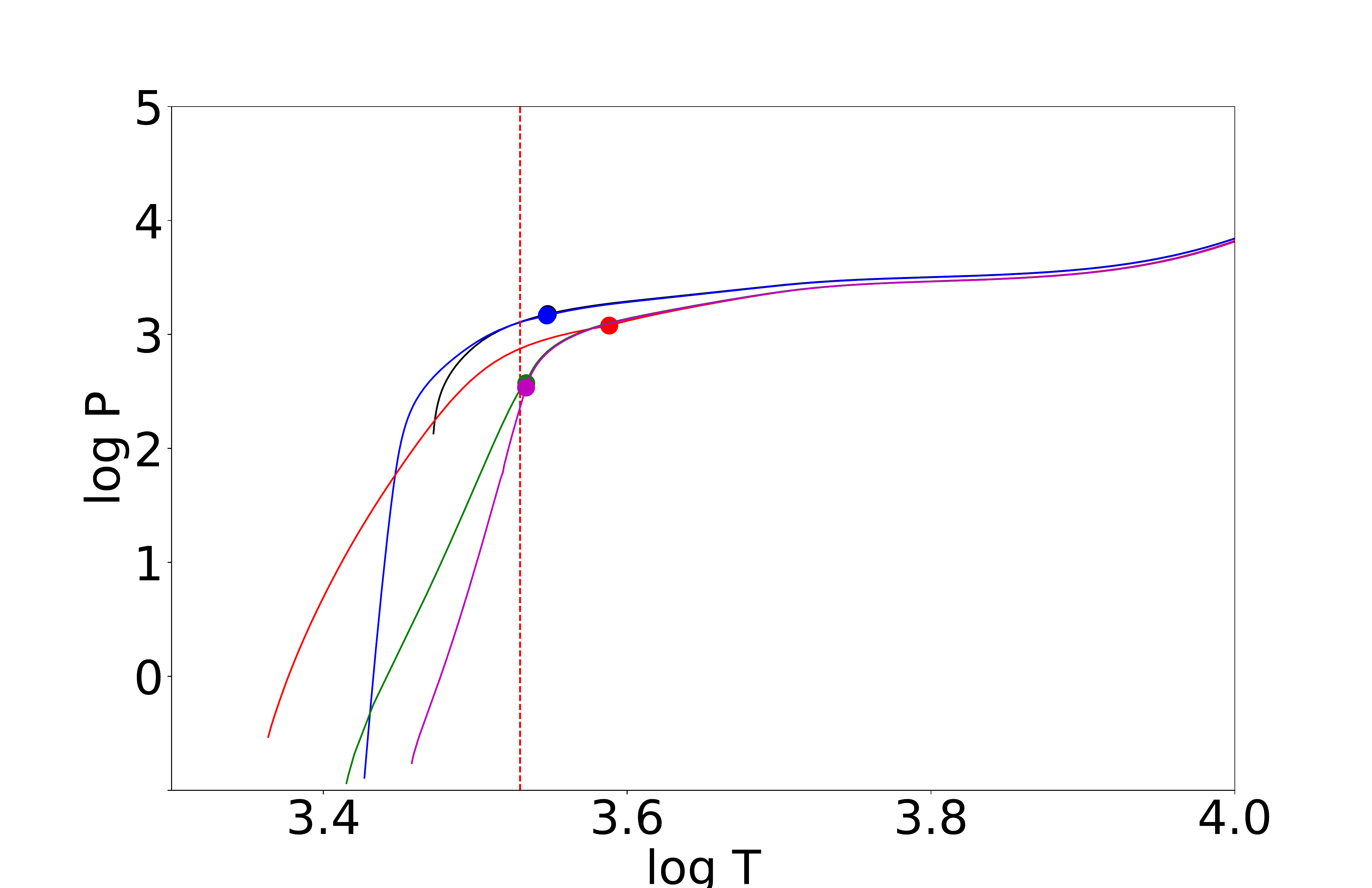}
   \caption{The pressure-temperature structure for the outer layers of the stellar model and atmosphere. Marker denotes position of BC, dashed line $T_{\rm eff}$ of COMARCS model. Black: pp-RM, blue: ss-RM, purple: pp-PM, green: ss-PM, red: COMARCS}
   \label{Fig: atmStruc}
\end{figure}

To address this question, Fig. \ref{Fig: atmStruc} shows $\log T$ vs $\log P$ for several atmospheric treatments, and the continuation into the interior stellar evolution model for a $3M_{\odot}$ model at the beginning of the TP-AGB phase with $\log(L/L_{\odot})=3.66-3.72$ and $\log T_{\rm eff}=3.53-3.55K$. The range in $L,T_{\rm eff}$ arises from all treatments being applied to the same interior model, without evolving the model, to understand the influence of the outer BC. In the case of the analytic treatments, pp/ss, the point at which the outer boundary condition is implemented corresponds to the effective temperature, and is marked by a black marker in each case. However, for the RT model, the effective temperature is defined separately and is marked by the vertical dashed line.

In all cases, the interior model converges to a similar solution, not far into the stellar interior. Although it is notable that the form of the RT model is not reproduced by the analytic models, $T_{\rm eff}$ is matched fairly closely for the PM opacities, as expected from the evolutionary tracks. Furthermore, the PM models follow the interior solution of the RT models from a shallower depth than the RM opacity models.

\begin{figure}
  \includegraphics[width=.9\columnwidth]{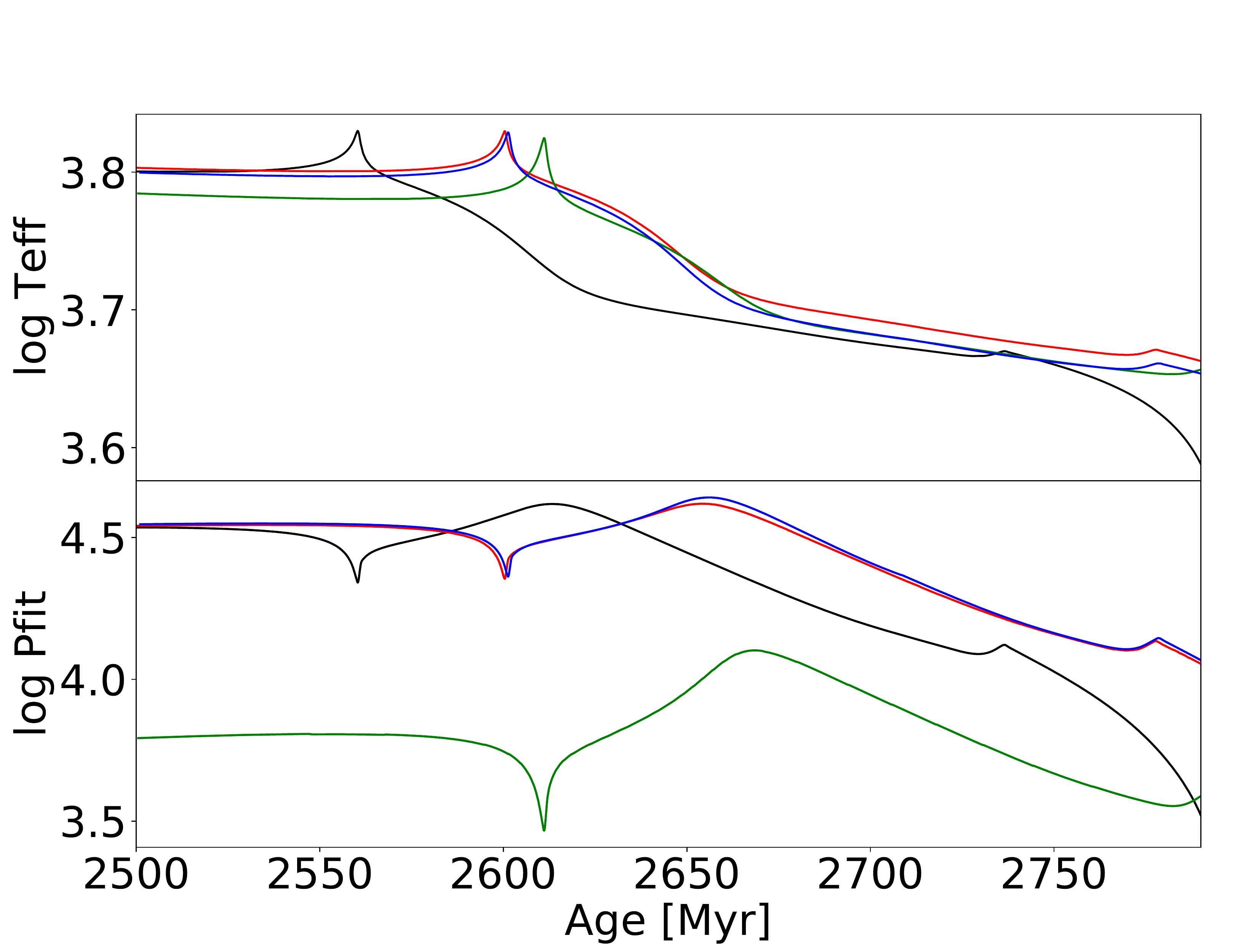}
   \caption{Effective temperature (top panel) and the pressure at fitting between interior and atmospheric model (lower panel) for the pre-AGB evolution of a 1.5$M_{\odot}$ model with boundary conditions represent denoted by Black: RM-pp, red: RM-sph, blue: RT (MARCS), green: PM-pp (all evolved with stated outer BC from ZAMS).}
   \label{Fig: preAGB Teff Pfit}
\end{figure}

Fig \ref{Fig: preAGB Teff Pfit} shows the effective temperature (top panel) and pressure at the fitting depth (lower panel) as a function of age for the same models as shown in Fig. \ref{Fig: hrd_preAGB}. This shows that physically, the PM opacities result in a significantly lower fitting pressure, and emphasises both the inability of the analytic approximations to replicate the physical properties of the RT models, and also to indicate how even this extreme change in the value of the pressure at the outer BC has such a negligible influence on the MS evolution. This is not to say that using the PM treatment is meaningless, as it has already been shown in Fig. \ref{Fig: atmStruc} that the interior model quickly forgets the outer BC, and that the PM appears to best reproduce the overall influence of the RT treatment.

Of course, it cannot be claimed that this is a fully justified physical description of the outer layers of the star, nevertheless, it does suggest that the initial belief that the opacity is likely to lie between the RM and PM is reasonably good. More importantly this provides a reasonable method for pursuing the full influence of this change to the outer boundary condition, which in some way mimics the behaviour of the RT models. Given this result, the PM-pp models were used to calculate a denser grid of models for additional study, and can be considered in some ways as a proxy for how the models may be changed by the inclusion of a full radiative transfer calculation for the atmosphere. The choice to continue using the pp models rather than ss model, coupled with the PM opacity, was taken as the opacity change appears to have a more significant influence than the geometry, which has little impact on the evolutionary models in the analytical framework investigated here. 

\subsection{Depth}

As has been previously mentioned, the equations of stellar structure which are implemented in the interior of the stellar evolution code, require that the diffusion approximation holds. Although this is commonly taken to be at a depth of $\tau =1$, this is not necessarily the case. As such it is worth considering the effect of attaching the grid of radiative transfer models at a greater depth.

Taking an evolutionary calculation for a $1.5M_{\odot}$ solar metallicity model, with a fitting depth of $\tau_{\rm fit}=1$, a separate RGB evolution with a fitting depth of $\tau_{\rm fit}=100$ was also calculated. The separate evolution for the $\tau_{\rm fit}=100$ model was only calculated once the star had begun the ascent of the RGB, so as to avoid the use of the plane-parallel models at higher log g, with the full evolution of the $\tau_{\rm fit}=1$ model using the MARCS grid of atmospheres. This was done for the sake of consistency as it is, at least in principle, necessary to account for the additional radial extent of the atmosphere between $\tau_{\rm fit}=1$ and $\tau_{\rm fit}=100$ in determining the outer BC.

\begin{figure}
  \includegraphics[width=.9\columnwidth]{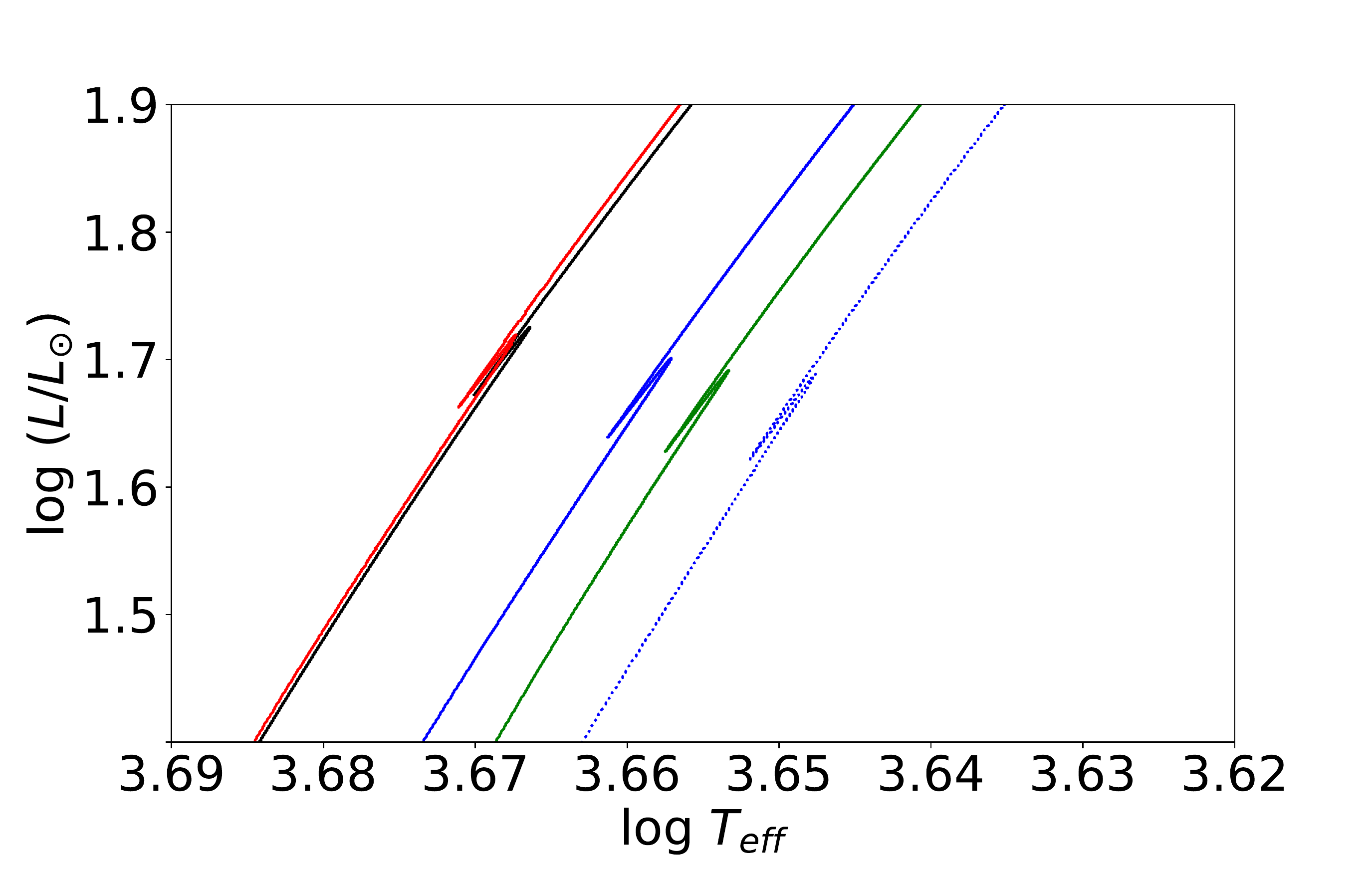}
   \caption{HR-diagram of the red bump for $1.5M_{\odot}$ models using the MARCS atmospheres, attached at $\tau_{\rm fit}=1$ (blue) and $\tau_{\rm fit}=100$ (blue-dotted). Other atmospheres also shown are black: RM-pp, red: RM-ss, green PM-pp (all evolved with stated outer BC from ZAMS).}
   \label{Fig: hrd_preAGB tau100}
\end{figure}

The resulting evolutionary tracks around the red bump on the RGB are shown in Fig. \ref{Fig: hrd_preAGB tau100} where the solid-blue line indicates the case where the MARCS atmospheres were attached at a depth of $\tau_{\rm fit}=1$ while the dotted-blue line shows a model where the fitting depth was shifted to $\tau_{\rm fit}=100$. Primarily, there is a clear shift in $T_{\rm eff}$, to lower temperatures, when attaching the model at greater depth.

Shown alongside these tracks are the same models which appear in Fig. \ref{Fig: hrd_preAGB}, where the black line is RM-pp, red is RM-ss and green is the PM-pp model. This shows that actually, attaching the MARCS atmospheres at a greater optical depht results in an effective temperature even lower than that given by the PM-pp models. Additionally, it can be seen that with the shift to lower temperatures, there is also a slight decrease in the luminosity of the red bump.

Typically, when attaching a model atmosphere to the interior stellar evolution code, the radius at the outermost point of the interior model is required in the calculation of log g, which is used for finding a corresponding solution from the atmospheric grid. In the case of the spherical models, there is a depth within the atmospheric model when the optical depth does not equal $\tau_{\rm fit}=1$. In principle, this radius must also be taken into account when using a grid of radiative transfer models as an outer BC, however, it is worth taking a moment to consider whether this is necessary.

\begin{figure}
  \includegraphics[width=.9\columnwidth]{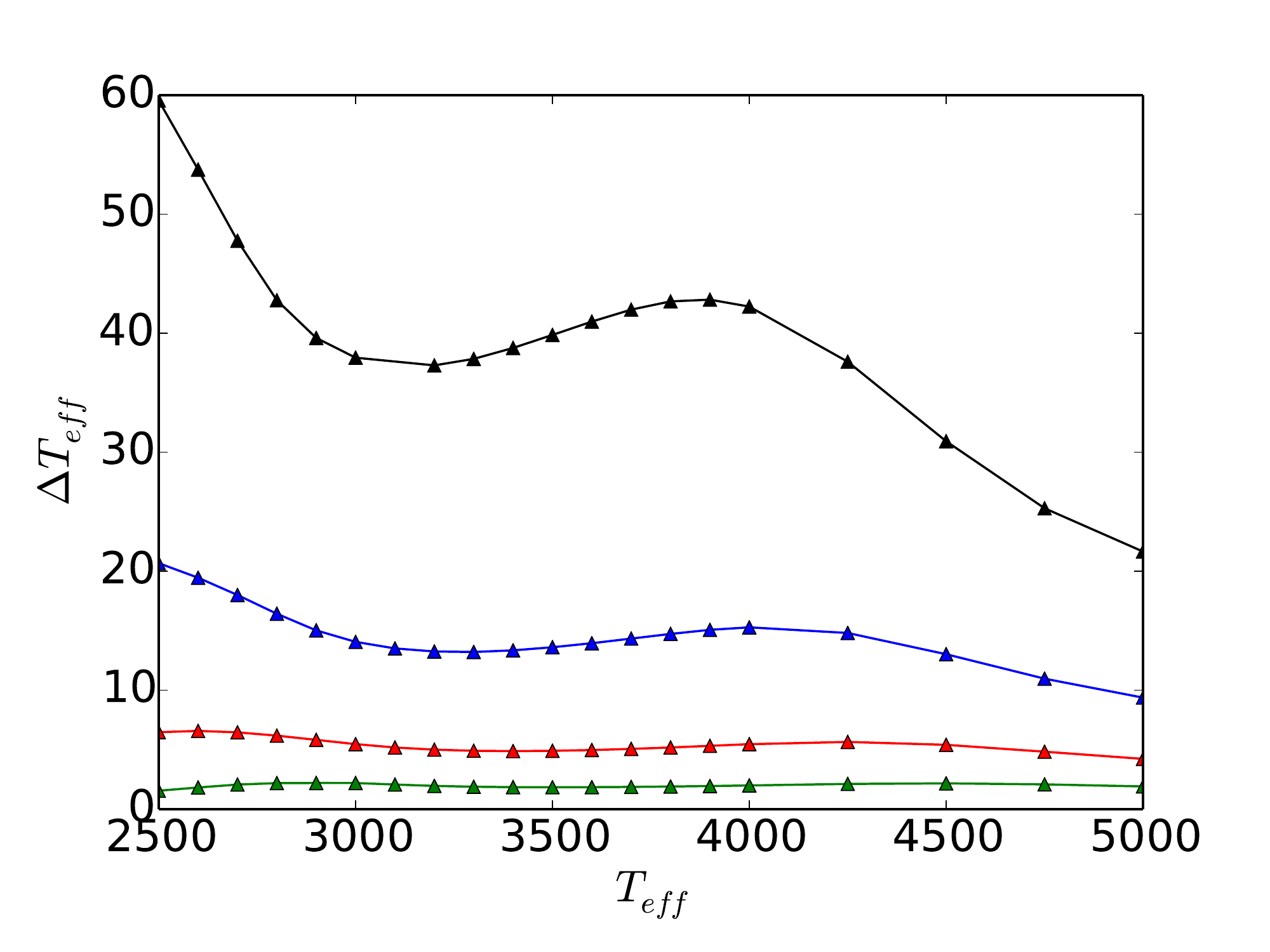}
   \caption{The resultant change in $T_{\rm eff}$ when radial extent of atmosphere is accounted for, when attaching a model at $\tau=100$. Shown as a function of $T_{\rm eff}$ for log g = 0 (black), 1 (blue), 2 (red), 3 (green) for a $1M_{\odot}$, solar metallicity MARCS atmosphere.}
   \label{Fig: delta Teff}
\end{figure}

As an estimate for the difference this change in radius might make to the outcome if neglected, it is possible to begin with the Stefan-Boltzmann law, $L = 4\pi R^2 \sigma_B T_{\rm eff}^4$. If it is assumed that the luminosity remains constant, then it can be said that a given radius corresponds to a particular effective temperature, and that given two models with different radii, $R_1$ and $R_2$, with the same luminosity and respective effective temperatures, $T_{\rm eff, 1}$ and $T_{\rm eff, 2}$, then the quantities in question can be related by $R_1^2 T_{\rm eff, 1}^4 = R_2^2 T_{\rm eff, 2}^4$, which means that based only on a change in the radius of a particular model, the simple relation
\begin{eqnarray}
 T_{\rm eff, 2}^4 = \frac{R_1^2}{R_2^2}T_{\rm eff, 1}^4
\end{eqnarray}
allows a new effective temperature to be determined.

Fig. \ref{Fig: delta Teff} shows the difference, as calculated in this way, for the change in the effective temperature for $1M_{\odot}$ MARCS atmospheric models, across a range of temperatures, with each line indicating a different log g (black: 0, blue: 1, red: 2, green: 3). It is not possible to calculate such a value for the plane-parallel models, given they have no atmospheric extent, however the fact that the difference for the higher log g values in the spherical models is so small is already an indication that the models are close to the plane-parallel approximation. On the other hand, for the low log g values, particularly towards the low-temperature regime, this is not the case. Indeed, it could be argued that in this regime, which concerns the TP-AGB, the extent of the atmosphere should also be included. 

This presents an additional problem, in that such depths are dependent on the mass of the model. Therefore, although it has been argued in Sect. \ref{Sec: RT} that it is justifiable to use only the $1M_{\odot}$ grid, due to the small structural differences between the different masses, this would be another reason for it to be preferable to use additional mass grids for consistency when considering fitting depths greater than $\tau=1$. For the further purposes of this work, a fitting depth of $\tau_{\rm fit}=1$ is always taken, although it should not be forgotten that the effective temperature would be lowered by fitting at a greater optical depth.

\subsection{Initial Impression}

On first viewing, the differences between atmospheric treatments are relatively minor, and with the numerical and physical uncertainties which are already known to influence the TP-AGB evolution (mass loss, mixing processes and nucleosynthesis to name a few. See \cite{Herwig2005} for more details),  it would be rather straightforward to dismiss such differences.

\begin{figure}
  \includegraphics[width=.9\columnwidth]{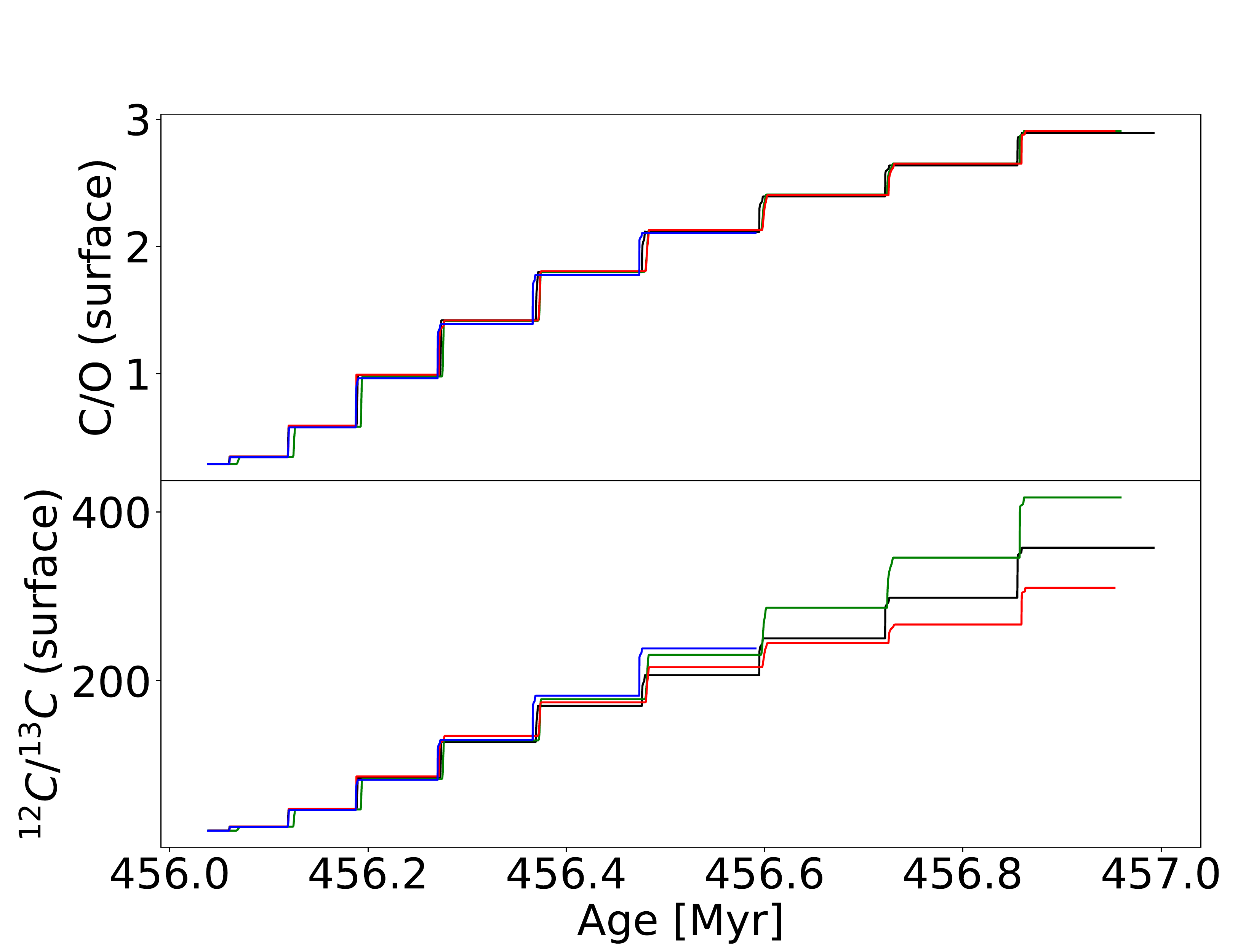}
   \caption{Surface C/O (upper panel) and $^{12}$C/$^{13}$C (lower panel) for the TP-AGB evolution of a $3M_{\odot}$ star. Outer BC altered such that black: RM-pp, red: RM-pp with C/O-interpolation, green: PM-pp, blue: RT (COMARCS)}
   \label{Fig: COratC12C13}
\end{figure}

The 3DU is a defining feature of the TP-AGB, so any change in the behaviour would be of great interest. Looking at the top panel in Fig. \ref{Fig: COratC12C13}, the changing C/O as a result of 3DU in a $3M_{\odot}$ can be seen. Although the overlap between models is not absolute, it would be difficult to justify any claim that the outer BC is playing any significant role on the 3DU and TP cycle based on looking at a single mass model. There is perhaps a slight change in the interpulse period, and a small difference in the C/O value, although nothing compared to the changes which are induced by variation of the overshoot parameter.

The lower panel of Fig. \ref{Fig: COratC12C13} shows the corresponding $\rm ^{12}C/^{13}C$ value. In this instance, the difference is more pronounced although the overall values are still of a comparable magnitude and nothing which could be distinguished in itself. 

\begin{figure}
  \includegraphics[width=.9\columnwidth]{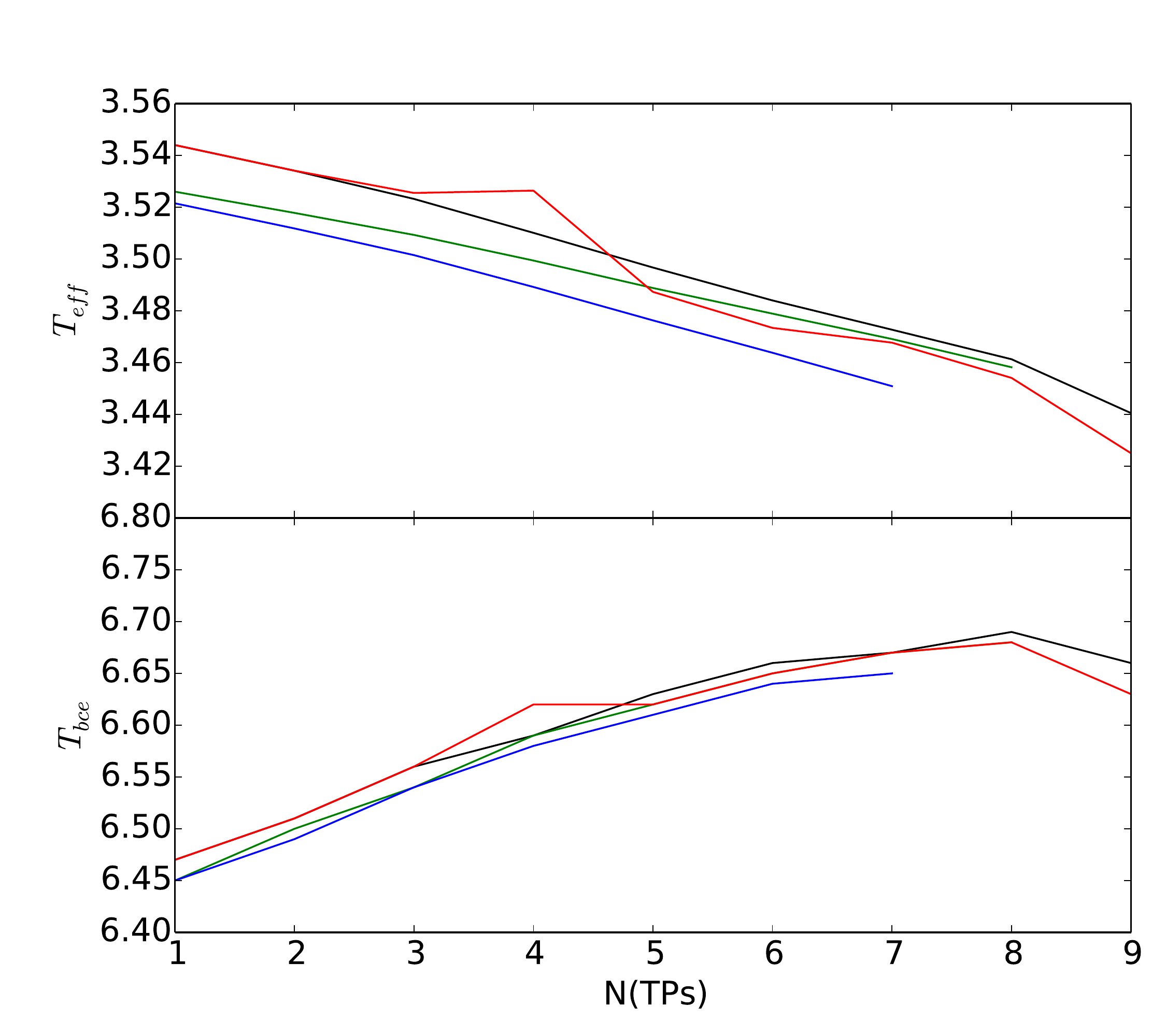}
  \caption{$T_{\rm eff}$ and $T_{bce}$ as a function of TP number for a $3M_{\odot}$ star. Outer BC altered such that black: RM-pp, red: RM-pp with C/O-interpolation, green: PM-pp, blue: RT (COMARCS)}

   \label{Fig: TeffTbce}
\end{figure}

Fig. \ref{Fig: TeffTbce} allows for a closer look at whether there is a resulting change in the interior of the stellar model. $T_{\rm eff}$ is shown in the top panel, while the temperature at the base of the convective envelope $T_{\rm bce}$ is shown in the lower panel, both as a function of TP number, where the temperature is taken at the minimum helium luminosity of the interpulse phase following a TP. This is done for clarity, and to hopefully allow for a more consistent representation of the differences since plotting the values as a function of time results in large variations over short timescales during the TP cycle.

As has already been seen in the previous section, the effective temperature of the PM and RT models are consistently lower than for the RM models. This does however change in the case of the RM-CO (red) model, where the variation of the opacity as a function of C/O results in an initially higher temperature with respect to the standard RM case, until it reaches the critical value of C/O=1. After this time, $T_{\rm eff}$ quickly begins to decrease to a value lower than the standard case, as is expected from other works \citep{Marigo2002,Weiss2009}.

The lower panel demonstrates that the change in the temperature at the surface is mirrored at least partially in the interior, and the depth to which the convective envelope descends. Taken in conjunction with the change in $\rm ^{12}C/^{13}C$ which is seen in the lower panel of Fig. \ref{Fig: COratC12C13}, it can be said that the outer BC is having at least some influence on the interior physics and as a consequence the 3DU is modified.

\subsection{Influence on Mass Loss}

\begin{figure}
  \includegraphics[width=.9\columnwidth]{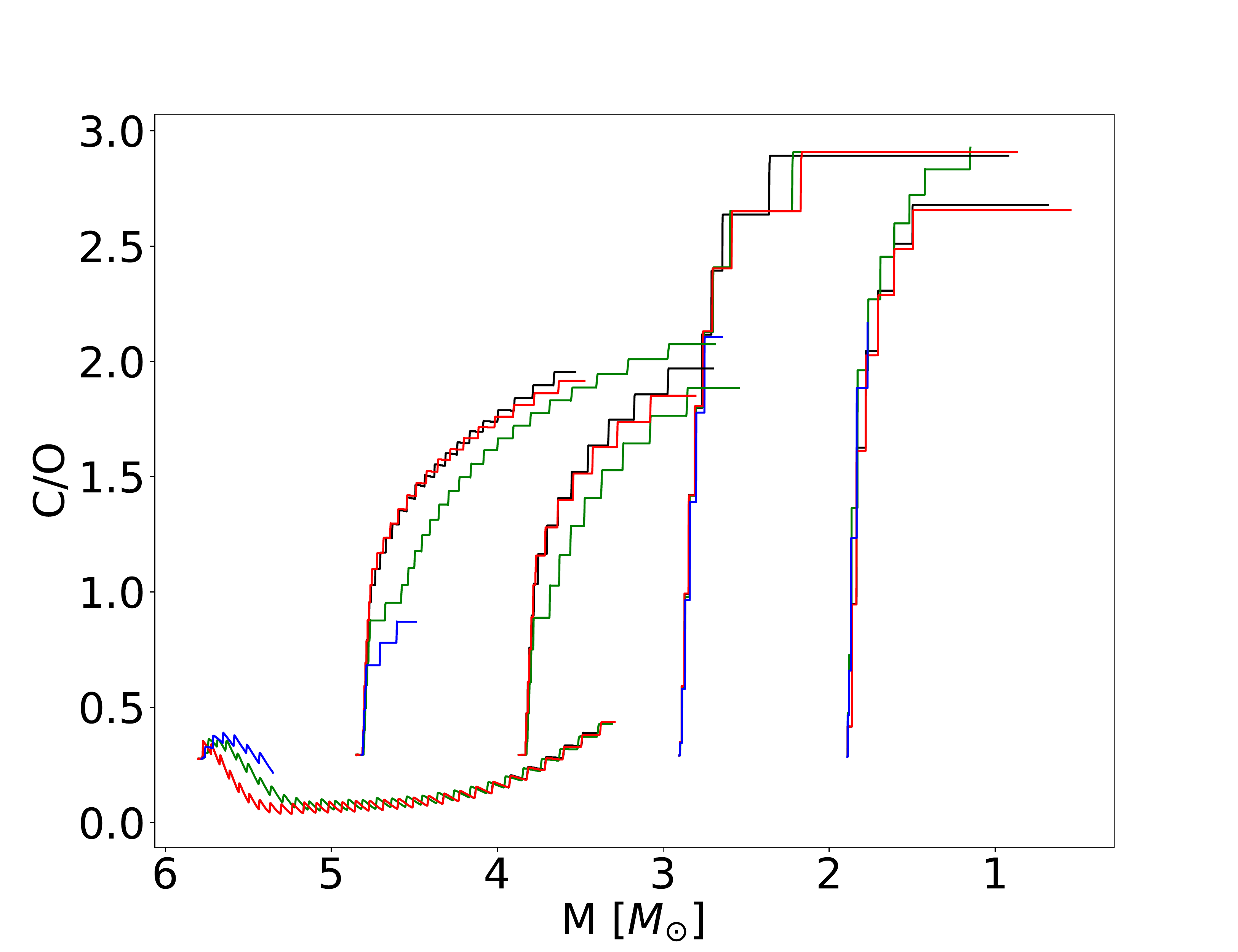}
   \caption{Carbon-to-Oxygen ratio as a function of mass during the TP-AGB evolution of models with ZAMS masses M=[2,3,4,5,6]. The colours denote different atmospheric boundary conditions given by Black: pp-RM, red: pp-RM-CO , green: pp-PM, blue: RT.}
   \label{Fig: COmass}
\end{figure}

A more direct result of altering the BC is the resulting mass loss history, which can be heavily dependent on $T_{\rm eff}$. Due to the lower effective temperatures, the PM and RT models start losing more mass earlier, so while the evolution can look fairly similar when considering the C/O value as a function of time, as in Fig \ref{Fig: COratC12C13}, it can be viewed somewhat differently when considered as a function of mass.

Fig \ref{Fig: COmass} shows the evolution of $2,3,4,5$ and $6 M_{\odot}$ mass models in the mass vs C/O plane, with the colours each representing a different BC. Especially in the case of the higher mass stars, a clear distinction can be seen between the various treatments, with RT and PM models having a substantially lower mass for the same abundance ratio. The evolution as a function of mass can also be considered an evolutionary indicator, given they are always losing mass.

In principle this would indicate a difference between models, however, it is only of a theoretical interest as it would not be possible to constrain masses and C/O values for these dust-enshrouded stars to a point such features could be distinguished. If that were the case, the mass loss history would be so tightly constrained as to provide empirical yield measurements and place far more observational constraints on the evolutionary models. However for a given mass loss description, the outer BC plays an important role, also influencing the yields. Along with being generally important for the chemical evolution of the galaxy, this is relevant for initial mass and thus timescale of second generation globular cluster stars, given the alternate composition and quantity of material expelled by the TP-AGB stars.

\subsection{Observable Quantities}

The results presented here focus on the PM-pp and RM-pp models, as this should allow for the extent of the outer BC influence to be investigated. The higher mass models ($\gtrsim 3M_{\odot}$) tend not to reach the final TP, and as such the lack of difference seen in the observable quantities is perhaps to be expected, as the strong mass loss which is the most significant outcome of changing the effective temperature is typically seen in the final stages of the TP-AGB evolution. All of the models in the mass range $1.6-2.8M_{\odot}$ have reached what is considered likely to be the final TP, i.e. it would be expected that if the calculation had continued to the post-AGB phase, it would not have required more TPs in order to lose any remaining envelope mass. In the case a model does not reach the final TP, no extrapolation has been carried out (for instance, as is done in ~\cite{Karakas2007}), as trying to determine when the final TP where significant mass loss occurs is fraught with danger. As such, these models are not discussed in as much detail, due to the greater uncertainties and the fact that large differences are in any case not seen.

\subsubsection{IFMR}


\begin{figure}
  \includegraphics[width=.9\columnwidth]{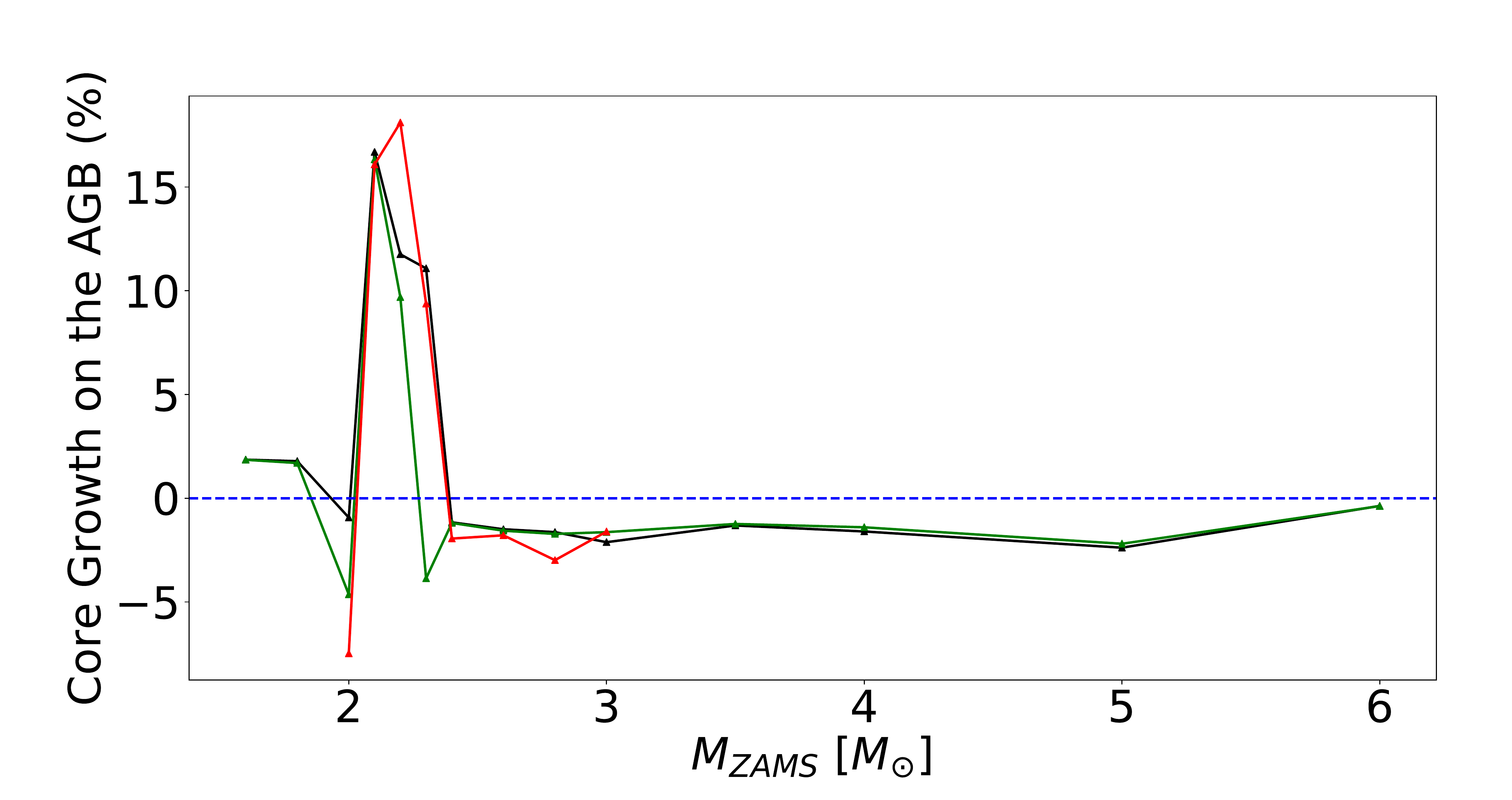}
   \caption{Percentage core growth during the TP-AGB for sets of models as denoted by black: RM-pp, green: PM-pp, red: PM-pp-ms}
   \label{Fig: growth}
\end{figure}

\begin{figure*}

  \makebox[\textwidth]{\includegraphics[width=0.9\paperwidth]{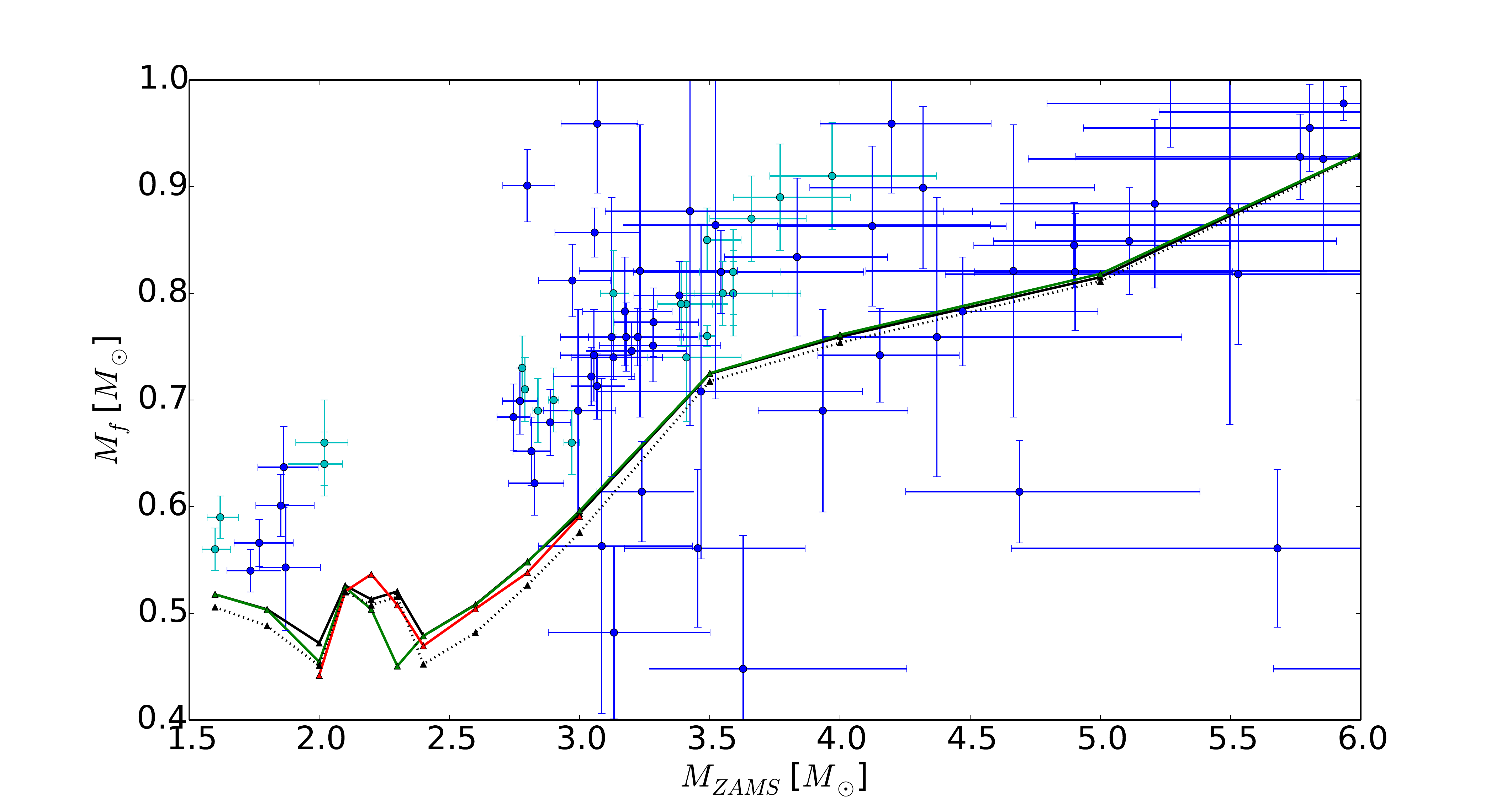}}

\caption{IFMR. Final masses denoted by solid line, dotted line discussed in text. Black: RM-pp, green: PM-pp, red: PM-pp-ms. Observational data points from open cluster white dwarfs. Observational data represented by blue makers ~\citep{Salaris2009} and cyan markers ~\citep{Kalirai2014}.}
\label{Fig: Plot IFMR}
\end{figure*}

\begin{figure}
  \includegraphics[width=.9\columnwidth]{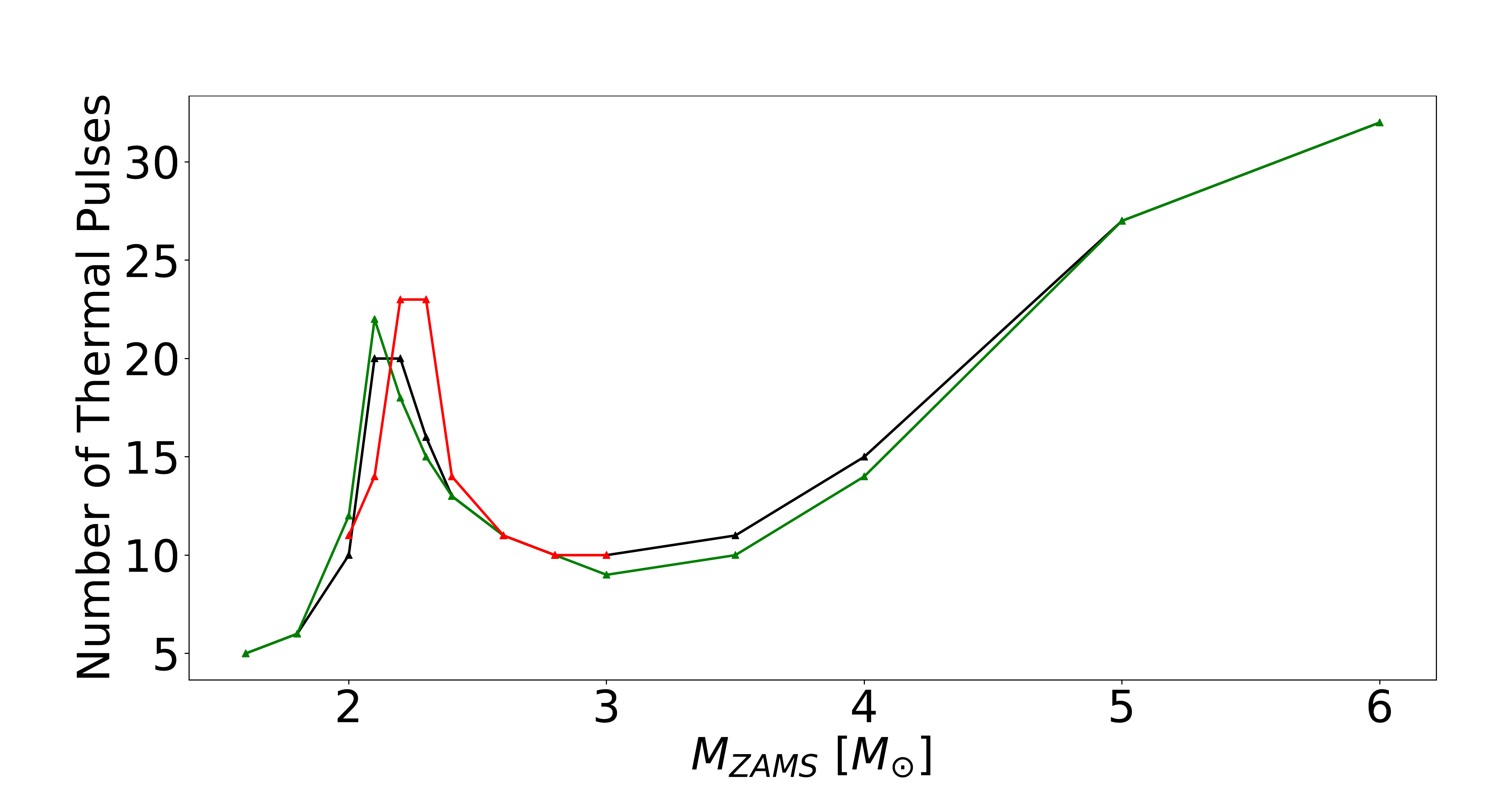}
   \caption{Number of thermal pulses undergone, as a function of ZAMS mass for models. Black: RM-pp, green: PM-pp, red: PM-pp-ms}
   \label{Fig: nTPs}
\end{figure}

The initial-final mass relation (IFMR) tests the integrated mass-loss of a star, and is one of the most concrete methods for testing TP-AGB stellar models, making it an obvious choice for investigating if the outer BC is having any systematic effect across a range of masses. This is particularly important given the sometimes chaotic nature of the TP-AGB evolution. 


The solid lines in Fig. \ref{Fig: Plot IFMR} indicate the IFMR for the RM-pp (black), PM-pp (green) and PM-pp-ms (red) treatments, with corresponding markers at the location of a model.  The data included in Fig. \ref{Fig: Plot IFMR} are observational constraints from open cluster white dwarfs, taken from Table 1 in ~\cite{Kalirai2014} and from the results of ~\cite{Salaris2009} where overshooting was included in the derivation. There is some overlap between the objects in each sample, but both are included as this also illustrates the uncertainty in the semi-empirical aspect of this relation. Additionally, the black dotted line shows the IFMR for the RM-pp models if the final mass is taken as the minimum value for the core mass from the last thermal pulse rather than the maximum value. This demonstrates the sensitivity of the results to the method of analysis.

It is already known ~\citep{Andrews2015} that there is some tension between the final masses of the models presented here and observations, however a comparison is shown nonetheless, as it is still beneficial to have some context when judging the changes to the models by changing the outer BC. The general trend seen in Fig. \ref{Fig: growth} is for the core growth of the models on the TP-AGB to be negative (excluding models in the range $2.1-2.3M_{\odot}$ for the moment). This is due to the treatment of overshooting, which is likely to be overly efficient during 3DU, resulting in a decreasing core mass as the model evolves. Therefore, rather than experiencing overall core mass growth during the TP-AGB, as would be expected \citep{Kalirai2014}, the models presented here can in fact see the core mass decrease due to the strength of the 3DU.

It can be seen in Fig. \ref{Fig: Plot IFMR} that there is negligible difference in the final core masses obtained as a result of changing the outer BC, which is surprising given the influence on the mass-loss which does result in a reduction in the number of thermal pulses, as can be seen in Fig. \ref{Fig: nTPs}, particularly in the range $2.4-4M_{\odot}$. This is surprising given the changes in the models which have been seen in previous sections, especially when considering the difference which could be obtained by defining the final core mass at a different phase of the TP cycle. However, there doesn't appear to be any doubt that changing the outer BC will not ease the tension of these models with the observations.

In the mass range $2.1-2.3M_{\odot}$, there is an increase in the core mass with respect to the general trend. This is due to the anomalous behaviour which is discussed further in \S \ref{Anom}. Clearly, if this behaviour were to be deemed to be physical, it is something which has the potential to have a significant impact on the observable quantaties.

\subsubsection{C-star Lifetime}
\begin{figure*}
  \makebox[\textwidth]{\includegraphics[width=0.9\paperwidth]{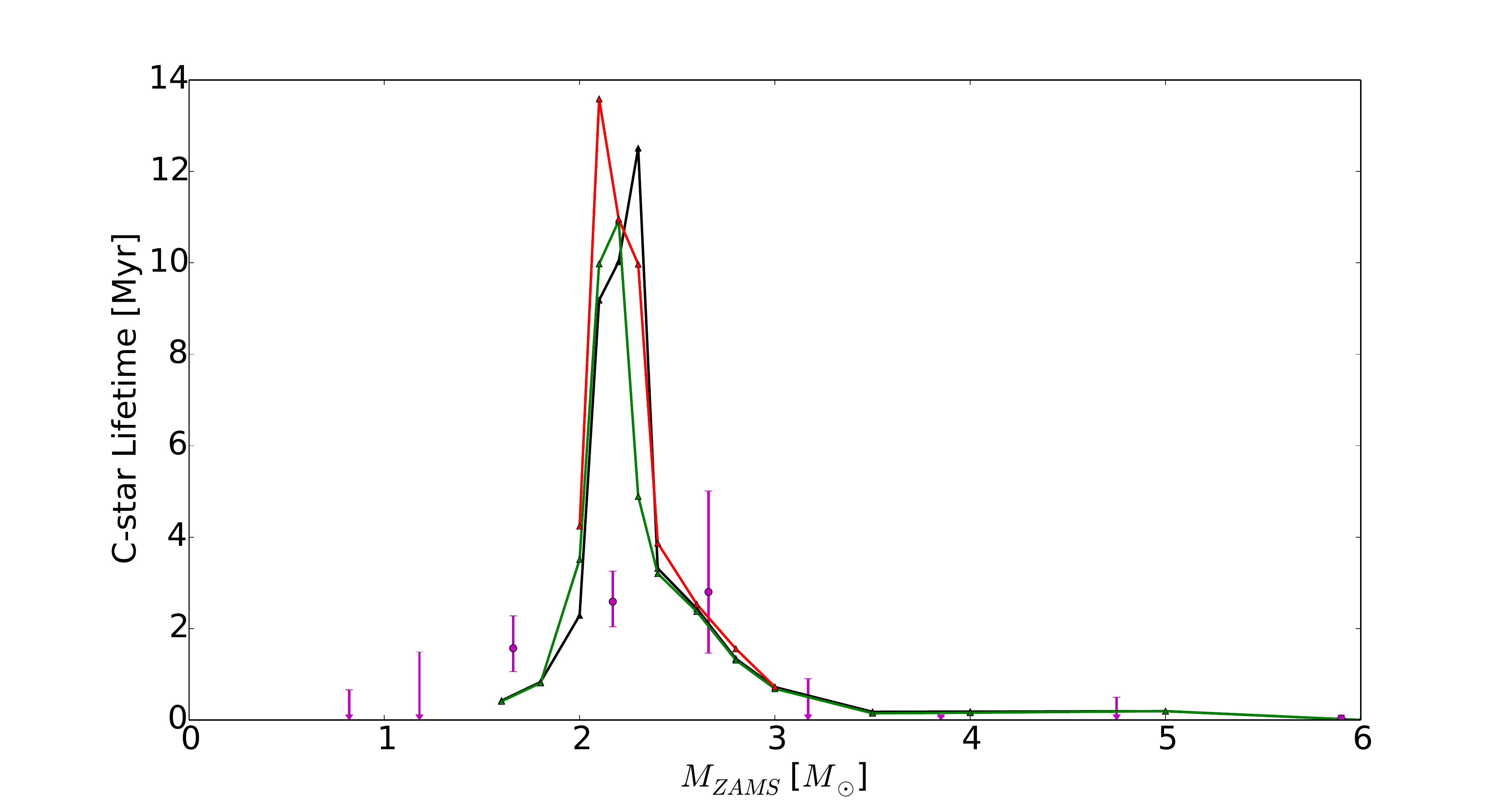}}
 \caption{Carbon-star lifetime for models denoted by black: RM-pp, green: PM-pp, red: PM-pp-ms. Observational markers taken from LMC data ~\citep{Girardi2007}.}
\label{Fig:  cLife}
\end{figure*}

The proportion of TP-AGB stars which are carbon-rich (C/O>1) allows for a useful diagnostic tool for this evolutionary phase. Fig. \ref{Fig: cLife} shows the time spent as carbon stars for the RM-pp, PM-pp and PM-pp-ms models, along with the data taken for the LMC from ~\cite{Girardi2007}. Although this data is for the LMC, this is commonly taken to be a metal fraction of Z=0.008, while the current work relies on solar metallicity models but with a value of Z=0.012, and is thus still a useful reference point.

It must again be taken into consideration, that the unexpected behaviour in the mass range $2.1-2.3M_{\odot}$ has a significant impact on the calculated quantities, and in this case the peak of the c-star lifetimes coincides with this region of initial masses. However, it is an explanation as to why the lifetimes presented here are a factor of 3 higher than might be expected. 

It can be seen that the values outside of this limited range are in line with expectation, and that there is again little to distinguish the change in the atmospheric treatment.

\subsection{Yields}

\begin{figure}
  \includegraphics[width=.9\columnwidth]{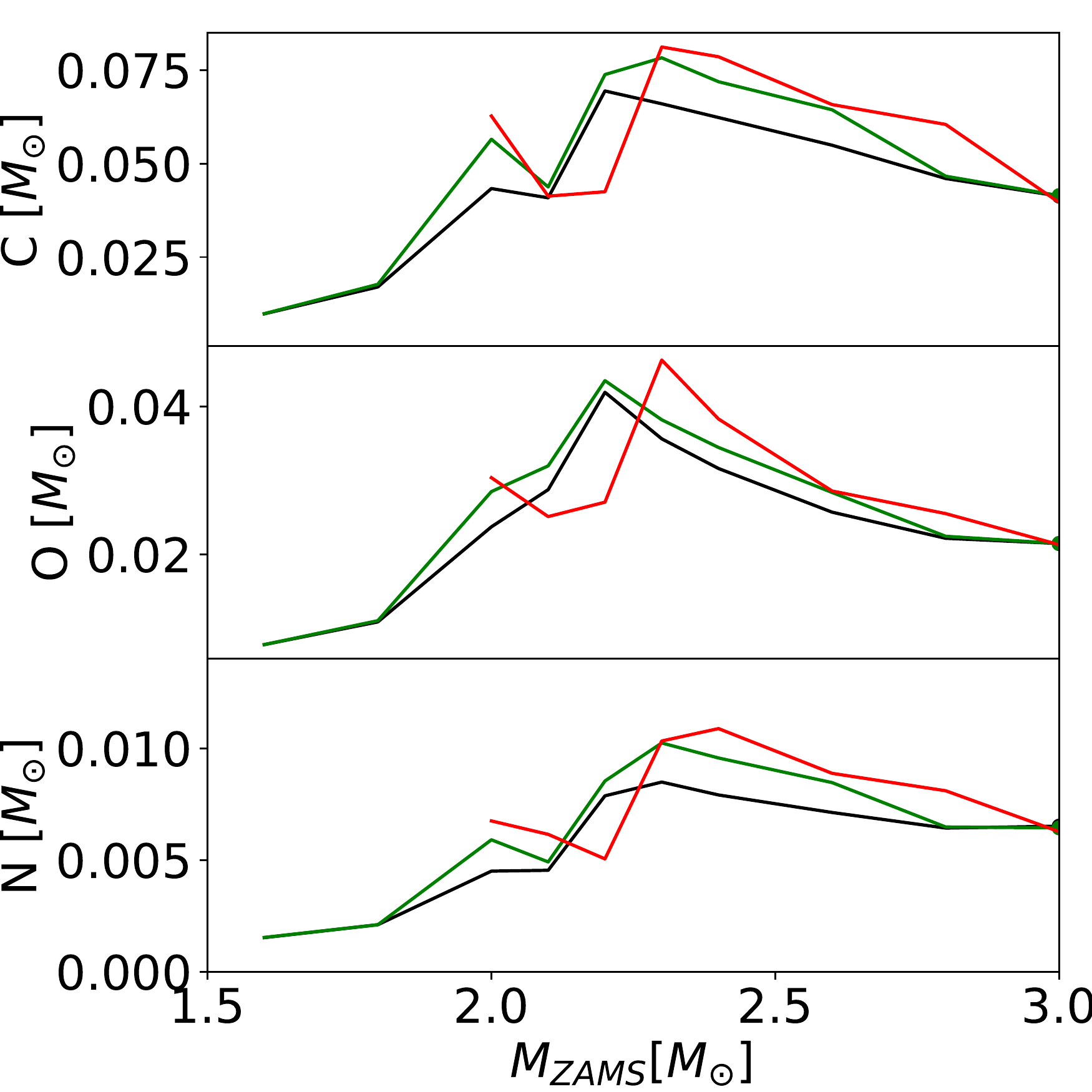}
   \caption{Total mass of Carbon, Oxygen and Nitrogen ejected into the ISM as a function of ZAMS mass for models with atmospheric boundary conditions denoted by Black: RM-pp, green: PM-pp, red: PM-pp-ms}
   \label{Fig: yields}
\end{figure}

For the wider astronomical community, TP-AGB models are of interest for their predictive power, and their contribution to the chemical evolution of host galaxies in particular. As such the C, N and O yields for the same models as in the previous section are presented as a function of ZAMS mass in Fig. \ref{Fig: yields}. s-process production elements would be highly interesting as a further test/exploration in this instance, given the connection to the base of the convective envelope which has been shown to be at least partially influenced by the outer BC, however, it is not considered here and must instead be left for future work. 

The clearest and most consistent change is in the range $2.4-2.8M_{\odot}$, with the standard RM-pp atmosphere less of all CNO elements. This change is likely due to the models including PM opacities becoming slightly more enriched from 3DU that the RM-pp case, such that when the majority of the envelope is expelled at the end of the star's TP-AGB evolution, the material contains more of the CNO elements


For masses $2.1-2.3M_{\odot}$, it is more difficult to say anything significant about the resulting yields due to the odd behaviour in the evolution, as discussed in the previous section. While for model with $<2M_{\odot}$, the lack of 3DU leaves the composition the same as at the first TP, so the same final mass for the models is then consistent with an equal chemical yield.

\subsection{Anomalous Behaviour}
\label{Anom}

\begin{figure}
  \includegraphics[width=\columnwidth]{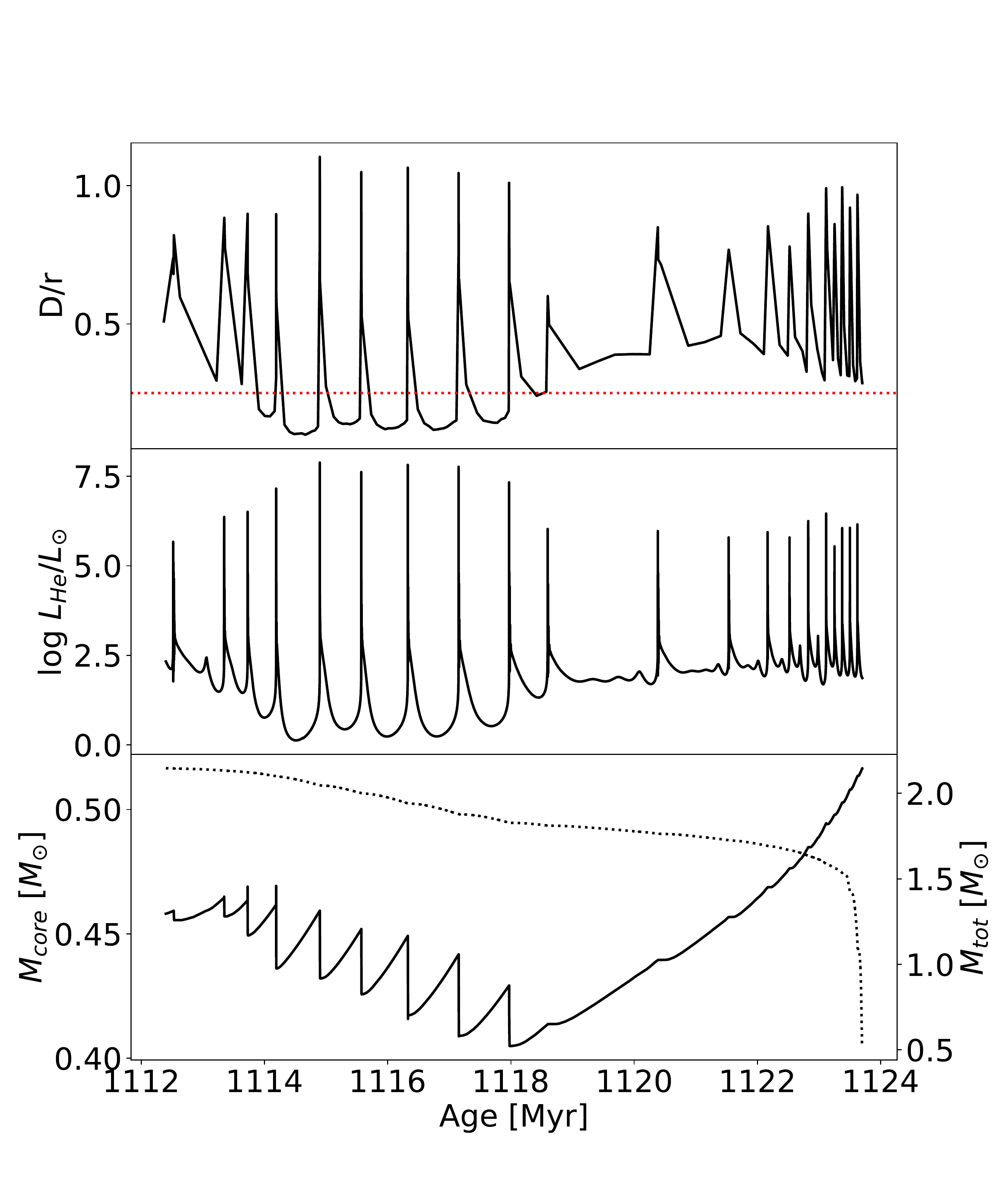}
  \caption[He-burning shell thickness over radius for a $2.2M_{\odot}$ model]{The ratio between the thickness of the helium burning shell and the radius at the base of the shell (top panel), the helium burning luminosity (middle panel) along with the core (solid line) and total mass (dotted line), both in the lower panel, for a $2.2M_{\odot}$ model. Horizontal red dotted line in the top panel indicates D/r=0.25.}
  \label{Fig: He shell thickness}
\end{figure}

Of particular note is the odd behaviour which was encountered in the models at a mass of $2.2M_{\odot}$, as seen in Fig. \ref{Fig: Plot IFMR}, which resulted in additional models being calculated at $2.1M_{\odot}$ and $2.3M_{\odot}$. Rather than an anomaly, a similar outcome is found with the addition of these masses, whereby the TP-AGB evolution is in one particular way quite different to what is observed at other masses. The middle panel in Fig. \ref{Fig: He shell thickness} shows the helium luminosity for a $2.2M_{\odot}$ star, where the unusual behaviour is quite apparent. In addition to the occasional irregular feature, the model progresses through the usual pattern of regular TPs, before experiencing an extended quiescent phase and then resuming thermally pulsing behaviour, but with a drastically reduced interpulse period and more erratic peak helium luminosity. During this secondary phase and beyond, 3DU does not occur, which leads to a continuing core growth, and hence the noticeably increased final masses in Fig. \ref{Fig: Plot IFMR}. This effect is likely due to these models being in the minimum core mass region, at the boundary between models which experienced a degenerate helium core flash at the end of the RGB phase and those which didn't, along with the strong 3DU resulting in an initially decreasing core mass, as seen in the lower panel of Fig. \ref{Fig: He shell thickness}.

In order to understand why the reduction in the core size appears to result in the suppression of the TPs, the top panel in Fig \ref{Fig: He shell thickness} shows the ratio of the shell thickness, D, to the radius at the base of the shell, r, for the same $2.2M_{\odot}$ model. The shell thickness and radius are only calculated when the full evolutionary model is stored, hence it is not as smooth a function as for the helium burning luminosity which is saved for every evolutionary time-step. The shell thickness is defined here by the region where the local helium burning exceeds 10 erg/g/s. This value was found to be low enough to ensure a helium burning shell is defined during the interpulse phase. The analytical argument outlined in \cite{Kippenhahn2012} requires that, for the thin shell condition, necessary for TPs, to be fulfilled, this ratio should not exceed $D/r=1/4$, for the case of an ideal monatomic gas. The quantitative value of this ratio is only included as a guideline, however, there is a notable increase in the value D/r as the model enters the quiescent phase before decreasing again as the TPs resume.

\begin{figure}
  \includegraphics[width=\columnwidth]{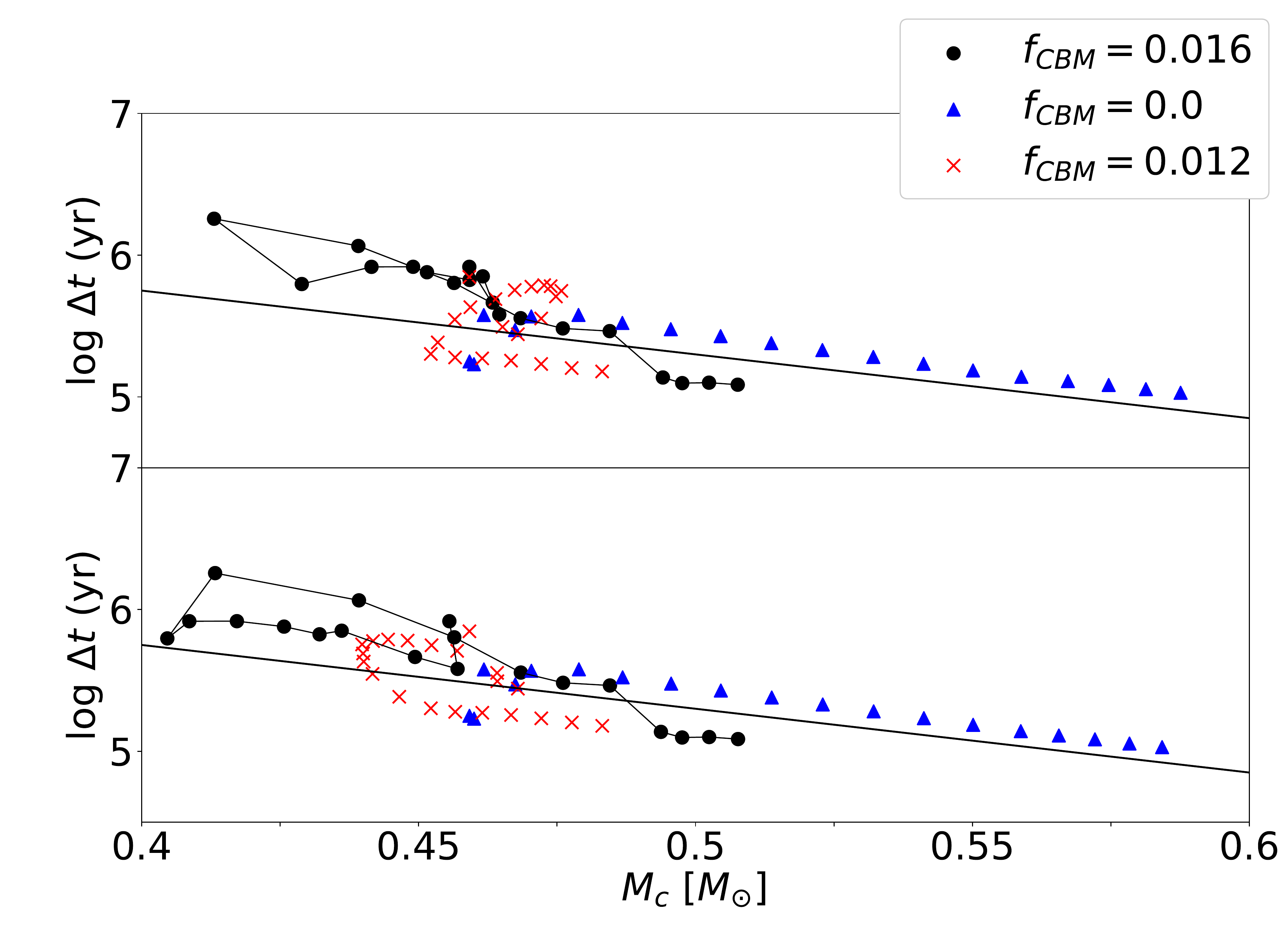}
    \caption{Core-mass vs interpulse period for models with varying overshoot parameter ($f_{\rm CBM}$) for a $2.2M_{\protect\odot}$ RM-pp model. Each marker represents the interpulse period with the core mass taken at the preceding thermal pulse. The top panel takes the core mass at its maximum before the thermal pulse, while the lower takes the core mass at its minimum after the thermal pulse.}
   \label{Fig: 2_2M Mc interpulse}
\end{figure}

A core mass-interpulse period relation \citep{Paczynski1975} has previously been empirically derived which helps to understand the reduction in the interpulse period after the quiescent phase, when the core begins to grow. For each thermal pulse of the models calculated, the core mass and subsequent interpulse period are shown in Fig. \ref{Fig: 2_2M Mc interpulse} with the top panel showing the maximum core mass prior to the interpulse period, the lower panel the minimum core mass. For the moment, only the lower panel is considered as the expectation would be this is what is directly related to the interpulse period, although the relation appears to hold in both cases. The solid line is the relation determined by \cite{Paczynski1975} and can be seen to be largely consistent with the models calculated here. Given the believed influence of the stong overshooting, additional models were calculated varying the mixing efficiency parameter $f_{\rm CBM}$, following the implementation of \cite{Herwig1997}. Markers are included for models which were evolved through the TP-AGB without any overshooting (blue triangles) and where a slightly reduced value for the mixing efficiency parameter, $f_{\rm CBM}$=0.012 (red crosses), is taken as compared to $f_{\rm CBM}$=0.016 (black circles, connected by a black line), the value for all other models presented within this work.

Beginning with the simplest case, where $f_{\rm CBM}$=0, it can be seen that after the first couple of thermal pulses, the increasing core mass and decreasing interpulse period agrees very well with the slope of the relation taken from \cite{Paczynski1975}, albeit shifted slightly above the line. A smooth behaviour is also seen in the case where $f_{\rm CBM}$=0.012, except that, to begin with, the markers move in the opposite direction due to the decreasing core mass, but similarly above the relation of \cite{Paczynski1975}. This model then proceeds to turn around as the third dredge-up is quenched and the core begins to grow, now running parallel to the relation but slightly below. This is also the observed behaviour for the initial $f_{\rm CBM}$=0.016 model, albeit with a jump due to the appearance of the quiescent phase. A similar quiescent phase was observed for models with $f_{\rm CBM}$=0.014 and $f_{\rm CBM}$=0.018, although for higher values the models did not converge. A condition which seems to indicate the presence of this quiescent phase is that the maximum core mass prior to the TP is $M_c\lesssim0.43M_{\odot}$.

This condition is met in part due to the decreasing core mass during the TP-AGB phase, however, it also appears to require the low core mass at the onset of the first TP. For comparison, initial 2$M_{\odot}$ models at Z=0.008 calculated with the Monash stellar code \citep[MONSTAR][]{Lattanzio1986,Campbell2008} and MESA \citep[version 8118]{Paxton2015} codes result in the core mass at the first thermal pulse being $\sim$0.55$M_{\odot}$ and $\sim$0.5$M_{\odot}$ respectively (Simon Campbell, private communication). It is also worth considering the sensitivity to the metallicity, for instance, core masses of 0.49$M_{\odot}$ at Z=0.02, 0.52$M_{\odot}$ at Z=0.01, 0.57$M_{\odot}$ at Z=0.001 all for 2$M_{\odot}$ models \citep{Bertolami2016}. These compare with a core mass of $\sim$0.47$M_{\odot}$ for a 2$M_{\odot}$ model calculated with GARSTEC at Z=0.008, which also experiences the anomalous behaviour, but which is not seen in any of the models run with other codes. However, it could be that physical treatment during an earlier evolutionary phase is leading to the anomalous behaviour by producing the lower core mass within GARSTEC in comparison to other codes. Although, it must be said that, given the spread in values from other codes and the variations with initial mass and metallicity, the core masses obtained with GARSTEC are not ridiculously low, although are certainly at the lower end. 



Overall, the evidence seems to be very much in favour of the anomalous behaviour being physical, as opposed to numerical, albeit something which is only possible under very specific conditions. However, it is seen across a range of masses with varying overshooting efficiencies during the TP-AGB and at different metallicities. It may only occur when a combination of factors culminate in the reduction in core size, from an already low value, where the models can also converge, yet it does appear to be physical. From the models considered here, a core mass below $\sim$0.43$M_{\odot}$ at the start of a thermal pulse is required to instigate this phenomenon.

Each step of the observed behaviour has an explanation in line with the expectations of the physical arguments. Initially, the strong overshooting causes the core mass to decrease, due to efficient 3DU, until the thin shell condition, requiring small D/r, is no longer fulfilled, quenching the thermal pulses. The lack of thermal pulses prevents the third dredge-up, allowing the core to grow until D/r is sufficiently small and the thermal pulses resume. As the core mass is now larger, the interpulse period is noticeably shorter than prior to the quiescent phase, in line with expectations from previously seen core mass-interpulse period relations \citep{Paczynski1975,Wagenhuber1998}. 

It should be noted that the effect likely only occurs in these models due to the overly-efficient 3DU implemented here, along with the low core mass at the first TP, as running the same model without overshooting avoids this behaviour altogether. This is not claimed to be a realistic or expected behaviour, but rather a physically consistent outcome for this mass-range based on the given set of input parameters and physics.



\section{Discussion/Future Work}
\label{Sec: Discussion}

The results presented in the previous section summarize the effects of implementing different atmospheric treatments in conjunction with a stellar evolution code. An initial point to note is that although the geometry has a minor influence on the problem, when comparing the plane-parallel and spherical analytic relations, it is not significant. In particular, changes resulting from this alteration are negligible in comparison to changing the mean opacity treatment used in conjunction with either the plane-parallel or spherical analytic method.

Indeed, changing the mean opacity used within the analytic frameworks appears to provide a method for anticipating the effect that using RT models for the full TP-AGB evolution would have, since it is not currently possible to do this directly. Section \ref{Sec: Plank Mean Proxy} outlines that using PM opacities in the outer layers of the atmosphere in some way mimics the behaviour of using the RT models as the outer BC. Although not ideal, this provides a platform to consider the implications for stellar evolution models based on current assumptions used in defining the stellar atmosphere.

Furthermore, it has been shown that attaching such models at greater optical depth moves the evolutionary tracks, at least on the RGB, to even lower effective temperatures than the use of the Planck mean opacity in the outermost layers. There is also a slight change in the luminosity of the red bump on the RGB in conjunction with this shift in effective temperature.



There is an effect on the evolution during the TP-AGB, primarily influencing the mass loss, resulting in changes to the models. However, the final outcome and observable predictions of the models remain largely unchanged for the cases considered here. It is surprising that the influence is so limited, but it would appear that, for the moment, other uncertainties within the TP-AGB models require more attention than the atmospheric treatment.


For the investigation to carry on further would require detailed RT models to be produced at very low log g values, in order to cover the entire evolutionary cycle of these models. In addition, it would be desirable to do so for a fine grid of C/O values, as it is clear this can rapidly change the outer BC in the region around C/O=1. More immediately it would be possible to further investigate whether the effects become stronger at different metallicities.


Additionally, an odd behaviour in models with a mass $2.1-2.3M_{\odot}$ was observed, believed to be related to these models having the lowest core mass at the beginning of the TP-AGB phase, carried over from the onset of helium core burning. This results in a prolonged and irregular TP-cycle, which even given the typically complex nature is somewhat extreme, with a quiescent phase followed by a new TP-cycle with a significantly reduced interpulse period, and a lack of 3DU. This behaviour has not previously been seen in TP-AGB models, however, it has been demonstrated that this is indeed likely to be a physical phenomenon rather than a numerical artifact, albeit one which only occurs under specific conditions which may themselves be unrealistic.


\section{Conclusion}
\label{Sec: Conclusion}

This work has considered the influence of the outer boundary condition on the evolution of TP-AGB stars, and presented the effects of changing standard treatments. The geometry and mean opacity treatment were investigated, along with a grid of radiative transfer models of stellar atmospheres.



There are changes to the evolution of the models as a result of changing the atmospheric treatment, however, there is negligible change in the final outcome and observable predictions of the models, such as the final core mass.  There is also a significant degeneracy with the mass loss prescription on the results of the model, and given the lack of an immediately realisable, physically realistic alternative for the outer BC, it is not currently necessary to consider the matter further. However, a final judgement on the matter would require full radiative transfer models covering the entire TP-AGB evolution, also covering both the C/O and mass parameter space.


A previously unseen behaviour was observed during the TP-AGB evolution, which appears to be physical rather than a numerical artifact, whereby the thermal pulses are briefly suppressed before resuming with a drastically reduced interpulse period. However, the circumstances for which it occurs may be unrealistic, or at least unlikely to appear. If it were present in stars, this would have a significant impact on the TP-AGB evolution and drastically change the observable quantaties in the narrow mass range it was observed, increasing the carbon-star lifetime above 10Myr and increasing the final core mass by $\sim0.1M_{\odot}$.

\section*{Acknowledgements}
This work was enhanced due to the support of the Deutscher Akademischer Austausch Dienst (DAAD), and the resulting discussions at Monash University. Particular thanks must also be given to Simon Campbell for helpful discussions and support. This work would not have been possible without the models provided by Bernhard Aringer and was furthered with the assistance of Gregory Feiden and others at Uppsala University. Finally, this work builds upon improvements made to the code during the Ph.D. Thesis of Agis Kitsikis, for which particular thanks is given.

%




\bibliographystyle{mnras}
\bibliography{ref} 






\bsp	
\label{lastpage}
\end{document}